\documentclass[usenatbib,usegraphicx]{mn2e}
\usepackage[fleqn]{amsmath}
\usepackage{aas_macros,amssymb,bm,charter,color,eulervm,multirow,array}
\newcommand{\e}[1]{\times 10^{#1}}                        

\newcommand{\eq}[1]{Equation \ref{eq:#1}}

\title[Analytical Hot Spot Shapes]{Analytical Hot Spot Shapes and Magnetospheric Radius from 3D Simulations of Magnetospheric Accretion}
\author[A. K. Kulkarni \& M. M. Romanova]{
A. K. Kulkarni\thanks{E-mail: akshay@astro.cornell.edu},
M. M. Romanova\thanks{E-mail: romanova@astro.cornell.edu} \\
Dept. of Astronomy, Cornell University, Ithaca, NY 14853
}

\begin{document}
\maketitle

\begin{abstract}

We present an analytical formula for the position and shape of the spots on the
surface of accreting magnetized stars in cases where a star has a dipole magnetic
field tilted at a small misalignment angle  $\Theta \lesssim 30^\circ$ about the
rotational axis, and the magnetosphere is 2.5-5 times the radius of the star.  We
observed that the azimuthal position of the spots varies significantly  when the
position of the inner disc varies.  In contrast, the polar position of the spots
varies only slightly because of the compression of the magnetosphere. The azimuthal
width of the spots strongly varies with $\Theta$:  spots have the shape of an arc at
larger misalignment angles, and resemble a ring at very small misalignment angles. The
polar width of the spots varies only slightly with changes in parameters. The motion
of the spots in the azimuthal direction can provide phase-shifts  in accreting
millisecond pulsars, and  the ``drift" of the period in Classical T Tauri stars. The
position and shape of the spots are determined by three parameters: misalignment angle
$\Theta$; normalized corotation radius, $r_c/R_\star$ and normalized magnetospheric
radius, $r_m/R_\star$, where $R_\star$ is the stellar radius.

  We also use our data to check the formula for the Alfv\'en
radius, where the main dependencies are $r_m\sim (\mu^2/{\dot M})^{2/7}$, where  $\mu$
is the magnetic moment of the star, and $\dot M$ is the accretion rate. We found that
the dependence is more gradual, $r_m\sim (\mu^2/{\dot M})^{1/5}$, which can be
explained by the compression of the magnetosphere by the disc matter and by the
non-dipole shape of the magnetic field lines of the external magnetosphere.

\end{abstract}



\section{Introduction}

Magnetospheric accretion occurs in a variety of astrophysical systems. The accreting
matter is stopped by the stellar magnetic field roughly at a distance from the star
where matter and magnetic stresses become equal. Beyond that point, matter flows
around the magnetosphere in a funnel flow and falls near the magnetic poles of the
star, forming hot spots (e.g., \citealt{GhoshLamb79}).

The light-curves observed from accreting magnetized stars are often associated with
the hot spots on their surfaces. In many applications it is important to know the
exact location and shape of the spots. For example, in millisecond pulsars, pulse
profiles are significantly affected by the hot spot location and shape (e.g.,
\citealt{PoutanenGierlinski03, IbragimovPoutanen09, LeahyEtAl09}). The position and
shape of the spots is also important for understanding the light-curves of Classical T
Tauri stars (CTTSs, e.g., \citealt{BouvierEtAl07}) and  in magnetized accreting white
dwarfs (e.g., \citealt{Warner95}).

Three-dimensional (3D) magneto-hydrodynamic (MHD) simulations of magnetospheric
accretion show that matter accretes to a star in two funnel streams, which form two
spots on its surface \citep{RomanovaEtAl03, RomanovaEtAl04, KulkarniRomanova05}.
Simulations show that the spots have the shape of a ring at small misalignment angles,
the shape of an arc at $\Theta\approx 30^\circ$ and the shape of a bar at very large
$\Theta$.  In all the cases, the energy flux is largest in the center of the spot and
gradually decreases outward.  The position of the spots does not coincide with the
magnetic pole.

The simulations also show that the funnel stream can be dragged by the disc so that
the spot forms at higher longitudes on the star's surface. Alternatively, it can trail
the magnetosphere, causing the spot to form at lower longitudes
\citep{RomanovaEtAl04}. Therefore, the azimuthal position varies with  accretion rate.
In the case of a small dipole misalignment angle, $\Theta\lesssim 5^\circ$, the funnel
stream may pass a whole cycle about the magnetic pole, so that a spot can move faster
or slower than the star (e.g., \citealt{KulkarniRomanova08, BachettiEtAl10}).

The motion of the spots along the stellar surface can produce observable effects such
as the phase shifts in light-curves, timing noise and intermittency of accreting
millisecond pulsars (e.g., \citealt{LambEtAl08, LambEtAl09, PatrunoEtAl09,
PoutanenEtAl09}). The motion of the spots can possibly explain the drifting periods
observed in many CTTSs (e.g., \citealt{RucinskiEtAl08}). Usually,  a simple model is
used for the spots, such as a circular spot centered at the magnetic pole, with a
constant or gaussian distribution of emitted flux.

More recent numerical simulations show that accreting magnetized stars may also be in
an unstable regime of accretion, where matter accretes  due to the
 magnetic Rayleigh-Taylor instability \citep{KulkarniRomanova08,
RomanovaEtAl08}. In this regime,  multiple chaotic `tongues'  penetrate through the
magnetosphere and form spots of chaotic shapes and positions. This regime is favorable
when the star rotates slower than the inner disc, though additional factors are
important for the onset of instability (e.g., \citealt{SpruitStehlePapaloizou95}). In
this paper, we only consider the set of parameters at which accretion is stable and
the spots have regular shapes.

The goal of this work is to derive a convenient formula for the shape and position of
the spots, and the dependencies of the parameters of the spots on the parameters of
the star and the disc.
 We concentrate on the cases of relatively small
misalignment angles, $\Theta\lesssim 30^\circ$, where the spots have the shape of an
arc or a ring.

We also use a set of our 3D MHD simulations to test the standard formula for the
Alfv\'en radius, which is routinely used for calculation of the truncation
(magnetospheric) radius, $r_m$. We found that the dependencies of  $r_m$ on the
magnetic moment of the star, $\mu$, and the accretion rate, $\dot M$,  are different
compared with the standard formula.

In Sec. 2 we describe our numerical model. In Sec. 3 we provide an analytical formula
for the spots. In Sec. 4 we investigate the dependence of the parameters of the spots
on the parameters of the star and the disc.  In Sec. 5 we provide a brief practical
guide on how to find the position of the spots from observation. In Sec. 6 we use our
simulations to test the formula for the Alfv\'en radius. In Sec. 7 we outline the main
conclusions from this work. Appendix A describes our reference values. Appendix B
shows parameters of the spots for all simulation runs. In Appendix C we derive the
formula for the magnetospheric radius and compare it with the standard formula for the
Alfv\'en radius.

\begin{figure*}
\centering
\includegraphics[width=8.0cm]{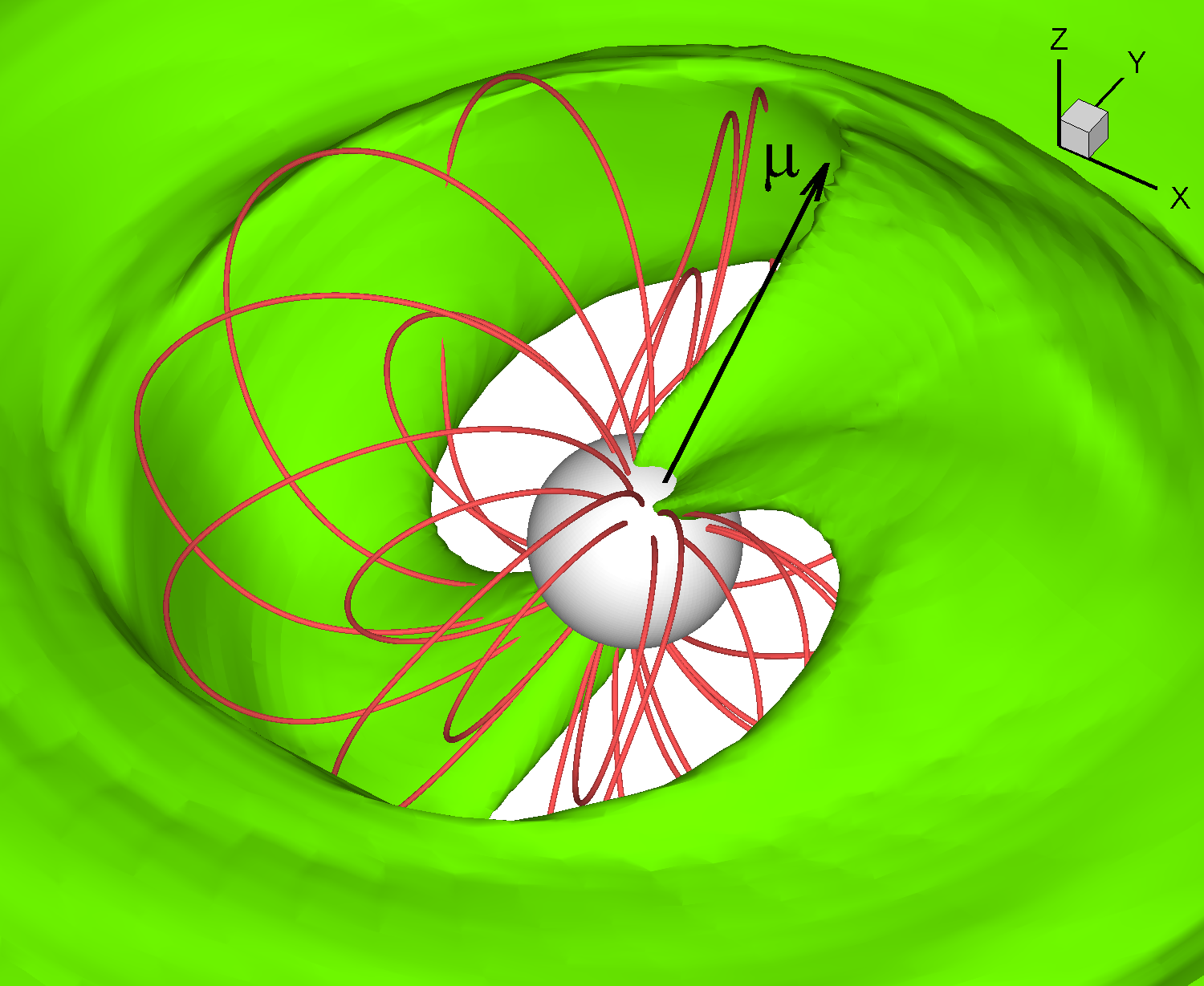}
\includegraphics[width=9.0cm]{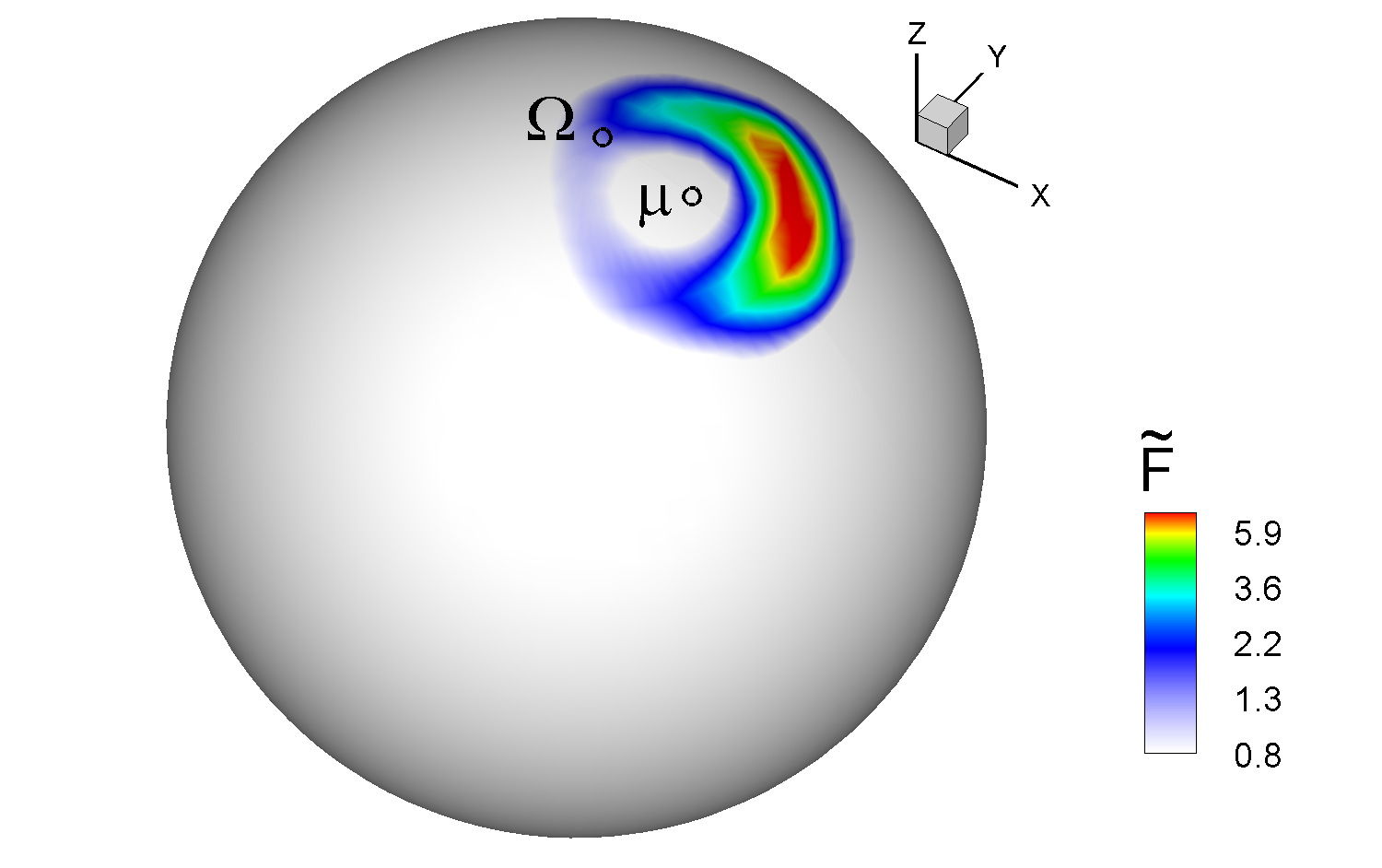}
\caption{\textit{Left panel:} A 3D view of the funnel flow from the disc to a
magnetized star, where the dipole moment $\mu$ is tilted by $\Theta=20^\circ$ about
the rotational axis.  One of the density levels is shown in green; sample field lines
are shown in red. \textit{Right panel:} the energy flux distribution on the surface of
the star. Circles show the position of the magnetic  ($\mu$) and rotational ($\Omega$)
axes, respectively. Other parameters are $\widetilde{\mu}=2$ and
$r_c=2$.}\label{3d-spot-2}
\end{figure*}

\section{Numerical model}
\label{sec:model}

To obtain the shapes of the spots, we perform global three-dimensional (3D)
magnetohydrodynamic (MHD) simulations of accretion from the disc onto a magnetized
rotating star.

 The model we use is the same as in our earlier 3D MHD simulations
(e.g., \citealt{RomanovaEtAl04, KulkarniRomanova08}). The star has a dipole magnetic
field, the axis of which makes angle $\Theta$ with the star's rotational axis. The
rotation axes of the star and the accretion disk are aligned. A magnetized star is
surrounded by an accretion disc which is cold and dense, and by a hot rarefied corona
which is 100 times less dense and 100 times hotter in the fiducial point.
  The disk and the corona are chosen to  be initially in a
quasi-equilibrium state, where the gravitational, centrifugal and pressure gradient
forces are in balance \citep{RomanovaEtAl02}. General relativistic effects, which are
important in neutron stars, are modelled using the Paczy\'nski-Wiita potential,
$\Phi(r) = GM_\star/(r-r_g)$ \citep{PaczynskiWiita80}, where $M_\star$ is the mass of
the star and $r_g \equiv 2GM_\star/c^2$ is Schwarzschild radius\footnote{Since the
simulations include general relativistic effects (approximated using the
Paczy\'nski-Wiita potential), strictly speaking they are only applicable to neutron
stars. However, previous studies \citep{KulkarniRomanova05} show that use of the
Paczy\'nski-Wiita potential does not affect the shape of the spots significantly. So
the results in this work can be applied to other types of accreting systems as well.}.
Viscosity is modelled using the $\alpha$-model \citep{ShakuraSunyaev73,
NovikovThorne73}. It is incorporated only in the disc and controls the accretion rate
through the disk. We take a small parameter $\alpha=0.02$ in all simulation runs.

To model accretion, the  MHD equations are solved numerically in three dimensions
using a Godunov-type numerical code, written in a ``cubed-sphere'' coordinate system
which rotates with the star \citep{KoldobaEtAl02}. The numerical approach is similar
to that described in \citet{PowellEtAl99}, where the eight-wave Roe-type approximate
Riemann solver is used to calculate flux densities between the cells.
 The grid resolution is identical to that in
\citet{KulkarniRomanova08} and equal to $N_r\times N^2=72\times 31^2$ in each of the
six blocks of the cubed sphere. Here, $N_r$ is the number of grids in the radial
direction, and $N$ is the number of grids in the angular directions in each of the six
sides of the cube.

 The boundary conditions at the star's surface amount to the assumption that the
infalling matter passes through the surface of the star, so that the dynamics of this
matter after it falls onto the star is ignored. At the external boundary, matter
inflow is permitted at the disc's part of the boundary, and outflow at the corona's
part of the boundary.

\begin{figure*}
\centering
\includegraphics[width=17cm]{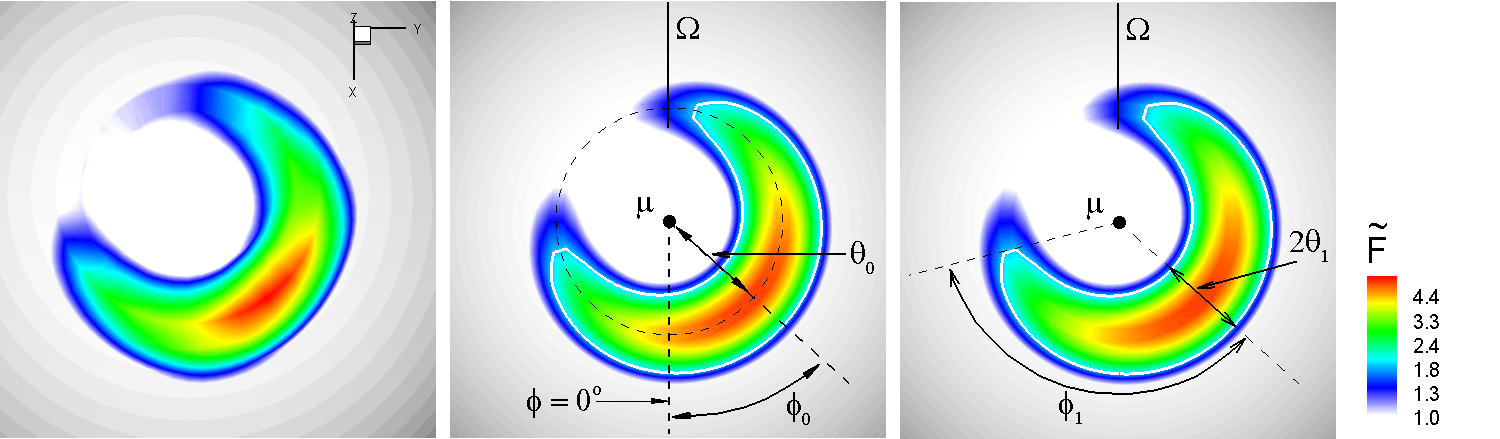}
\caption{\textit{Left panel:} Energy flux distribution on the surface of the star
obtained in one of the simulation runs, with parameters  $\Theta=10^\circ,
\widetilde{\mu}=1.5$, $r_c=1.8$, at time $t=8$. {\it Middle panel:}  Normalized flux
distribution in the fitted spot, calculated with Eq. \ref{eq:spotflux}. The white line
shows the boundary where the flux is $e-$times smaller than in the center of the spot.
The position of the spot in polar, $\theta_0$, and azimuthal, $\phi_0$, directions is
shown.  \textit{Right panel:} The same as the middle panel, except that the width of
the spot is defined in polar, $\theta_1$, and azimuthal, $\phi_1$, directions.}
\label{spotfit-3}
\end{figure*}

The simulations are done using dimensionless variables. The dimensionless value of
every physical quantity $q$ is defined as $\tilde{q} = q/q_0$, where $q_0$ is the
reference value for $q$. Appendix \ref{app:refval} shows how the reference values are
determined, and lists the reference values for three classes of central objects:
classical T Tauri stars, white dwarfs and neutron stars. Subsequently, we drop the
tildes above the dimensionless variables and show dimensionless values everywhere
unless otherwise stated. For clarity, we keep the tildes above the dimensionless
magnetic moment of the star, $\widetilde\mu$,  and the matter and angular momentum
fluxes $\widetilde{\dot M}$ and $\widetilde{\dot L}$, respectively.

 Fig. \ref{3d-spot-2} (left panel) shows an example
of funnel streams obtained in a typical 3D MHD simulation run. The right panel shows
the energy flux distribution at the surface of the star (the hot spot). One can see
that the spot has the shape of an arc, and the center of the spot (the highest energy
flux) is off-set from the magnetic pole. The center of the spot is also off-set from
the $\mu-\Omega$ plane, because the inner disc rotates faster than the magnetosphere,
and both the funnel stream and the spot are shifted. In this paper, we provide an
analytical formula for the shapes and positions of such spots.

\section{Analytical Formula for the Spots}
\label{sec:analytical spot shape}

For relatively small misalignment angles, $\Theta\lesssim 30^\circ$, the position and
shape of the spots can be approximated by analytical formula. Namely, each antipodal
spot can be well approximated as a circular arc centered at the magnetic pole, with a
gaussian flux distribution centered some distance away from the magnetic pole, as
follows (see Fig. \ref{spotfit-3}, two right panels):
\begin{equation}
\label{eq:spotflux} F(\theta,\phi) = F_c exp \left\{ - \left[ \left(
\frac{\theta-\theta_0}{\theta_1} \right)^2 + \left( \frac{\phi-\phi_0}{\phi_1}
\right)^2 \right] \right\}.
\end{equation}
The spherical polar and azimuthal angles $\theta$ and $\phi$ are measured with respect
to the magnetic axis $\mu$. The azimuthal position $\phi=0$ is defined by the
$\mu-\Omega$ plane. The spot is described by the following parameters:

\begin{itemize}

\item The polar and azimuthal position of the spot $(\theta_0, \phi_0)$.

\item The polar and azimuthal width of the spot $(\theta_1, \phi_1)$, which are
determined to be a half-width of the gaussian in the polar and azimuthal directions.
The half-width is defined as the distance from the center of the spot, to the radius
at which the emitted flux is 1/e times smaller than at the spot center.


\item  The flux  emitted from the center of the spot, $F_c$.

\end{itemize}

We then apply a least-squares fit of this expression (given by Eq. \ref{eq:spotflux})
to the spots obtained from the MHD simulations. Fig. \ref{spotfit-3} compares the
northern hot spot from the simulation, where misalignment angle $\Theta=10^\circ$
(left panel), with the fitted spot (right two panels). The fitted spot agrees well
with the simulation. This procedure, when applied to the spots obtained at different
parameters, shows that the rms error of the approximation is usually a few per cent.

\eq{spotflux} gives the flux distribution for the northern hot spot. The southern hot
spot is generally identical to the northern one for a dipole field,  so that
parameters describing it are simply $(\pi-\theta_0, \pi+\phi_0)$ and
($\theta_1,\phi_1)$. When using Eq. \ref{eq:spotflux} with these parameters, care must
be taken to ensure that ${|\phi-\phi_0| \leq \pi}$ (by adding integral multiples of
$\pi$ to $\phi$) for northern spots and $|\phi-(\pi+\phi_0)| \leq \pi$ for southern
spots.

\section{Dependence of spot parameters on parameters of the star}
\label{sec:results}

We performed a large set of 3D MHD numerical simulations at different misalignment
angles of the dipole moment $\Theta$,   periods of the star $P_\star$ and its magnetic
moment $\widetilde{\mu}$. The disc is the same in all simulation runs.  Our
simulations and analysis show that the location and shape of the spots are determined
by the following parameters:

\begin{enumerate}

\item Misalignment angle $\Theta$. We varied this parameter from very small values up
to relatively large values:  $\Theta=2^\circ, 5^\circ, 10^\circ, 20^\circ, 30^\circ$.

\item  Magnetospheric radius $r_m$, normalized by stellar radius $r_m/R_\star$. The
magnetospheric radius can be determined in several ways. For example, an equality of
matter and magnetic stresses, $\beta_1=(p+\rho v_\phi^2)/(B^2/8\pi)=1$, usually
corresponds to the innermost region of the disc, where the disc density decreases
sharply (e.g., \citealt{RomanovaEtAl11}). On the other hand, the condition $\beta=8\pi
p/B^2$ often corresponds to the region in the disc where the funnel flow begins
\citep{BessolazEtAl08}. Our simulations show that the condition $\beta=1$ corresponds
to the radius from which matter flows along a funnel stream to the central regions of
the spots. This is why we chose the condition $\beta=1$ to derive $r_m$.  We varied
the size of the magnetosphere using the dimensionless magnetic moment of the star
$\widetilde{\mu}$ in the range of $\widetilde{\mu}=0.5-2$, and obtained the
magnetospheric radius in the range of  $r_m/R_\star\approx 2.4-4.9$.

\item  Dimensionless period of the star $P_\star$, or the ratio $r_c/R_\star$, where
$r_c=[GM_\star (P_\star/2\pi)^2]^{1/3}$ is the corotation radius (here $P_\star$ is
the dimensional period). In our model, we  use either dimensionless period $P_\star$,
or the ratio $r_c/R_\star$. We chose three periods for the star: $P_\star=1.8, 2.4,
2.8$ which correspond to $r_c/R_\star=4.2, 5.1, 5.3$ (see Table \ref{tab:P-rc} for the
conversion between dimensional and dimensionless values).

\end{enumerate}

\begin{figure*}
\centering
\includegraphics[width=7.5cm]{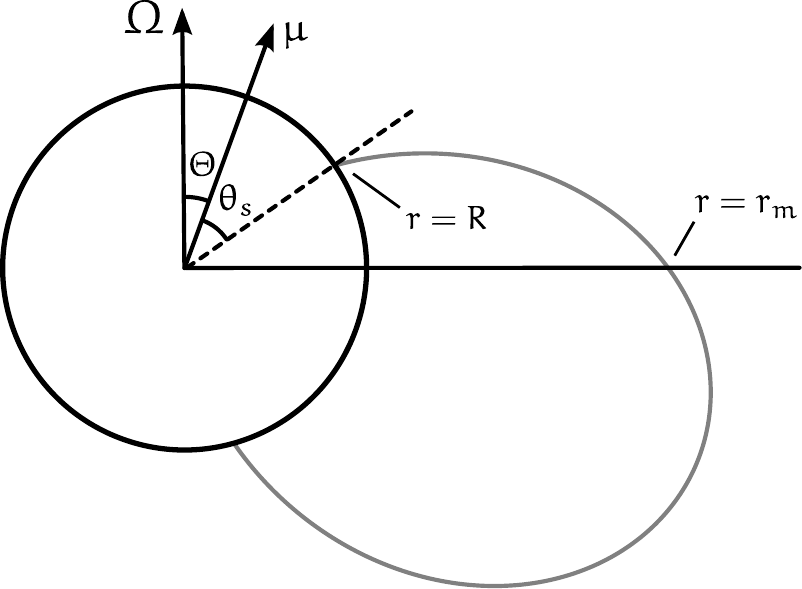}
\includegraphics[width=9.cm]{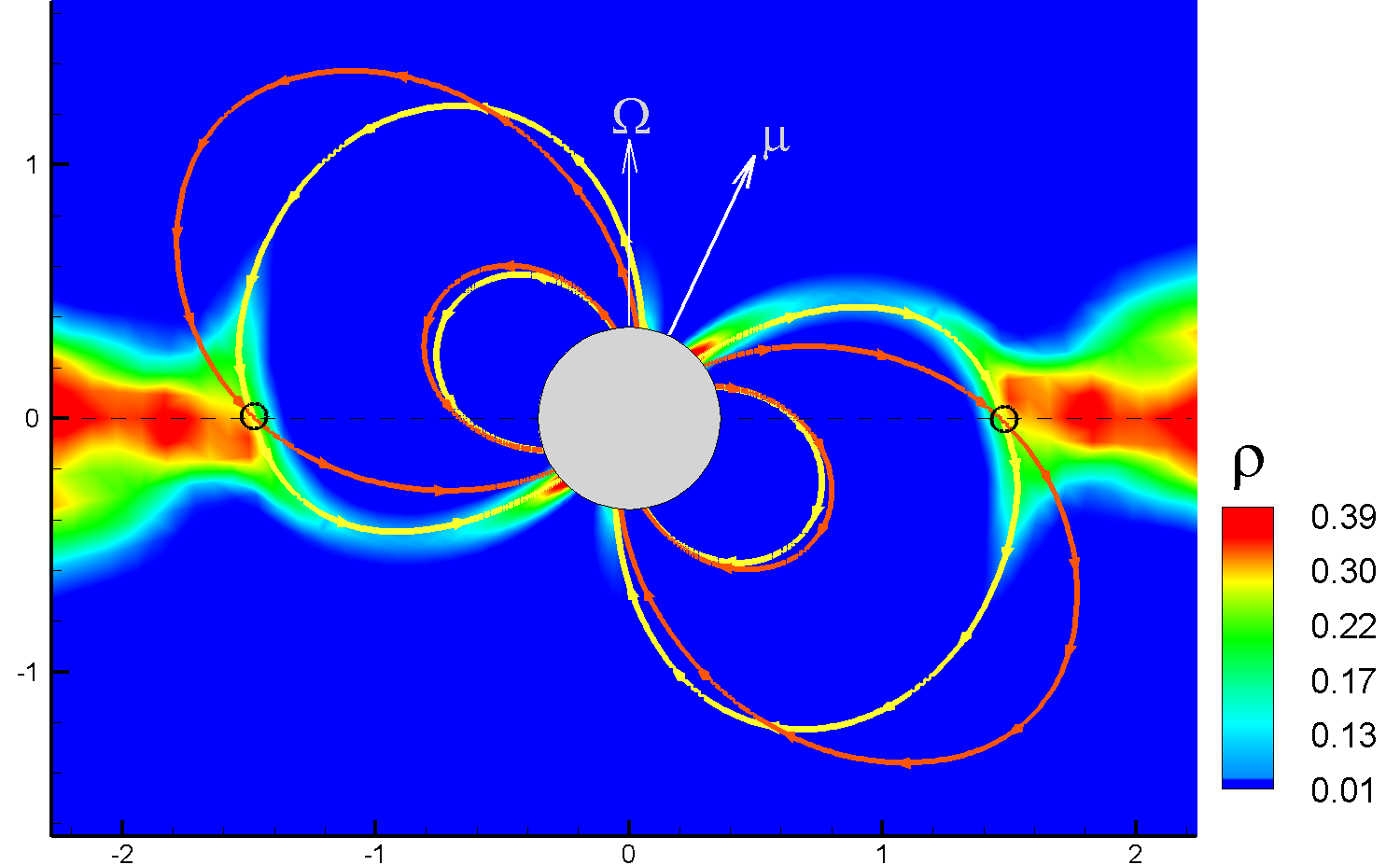}
\caption{\textit{Left Panel:} The sketch shows a star with a dipole magnetic field,
the position of the inner disc $r=r_m$ and the polar angle $\theta_s$ of the spot's
position on the star. The magnetic axis $\mu$ of the dipole moment is tilted at an
angle $\Theta$ about the rotational axis $\Omega$. \textit{Right Panel:} Density
distribution (color background) and magnetic field lines (yellow lines) in a typical
simulation run ($\Theta=30^\circ$, $r_c=1.8$, $\widetilde{\mu}=1.5$, at $t=15$). Red
lines show the dipole field lines at $t=0$. The external field lines are chosen such
that both the yellow and the red lines start at the inner magnetospheric radius $r_m$,
which is marked as a black circle.}\label{dipole-spot}
\end{figure*}

\begin{figure*}
\centering
\includegraphics[width=16cm]{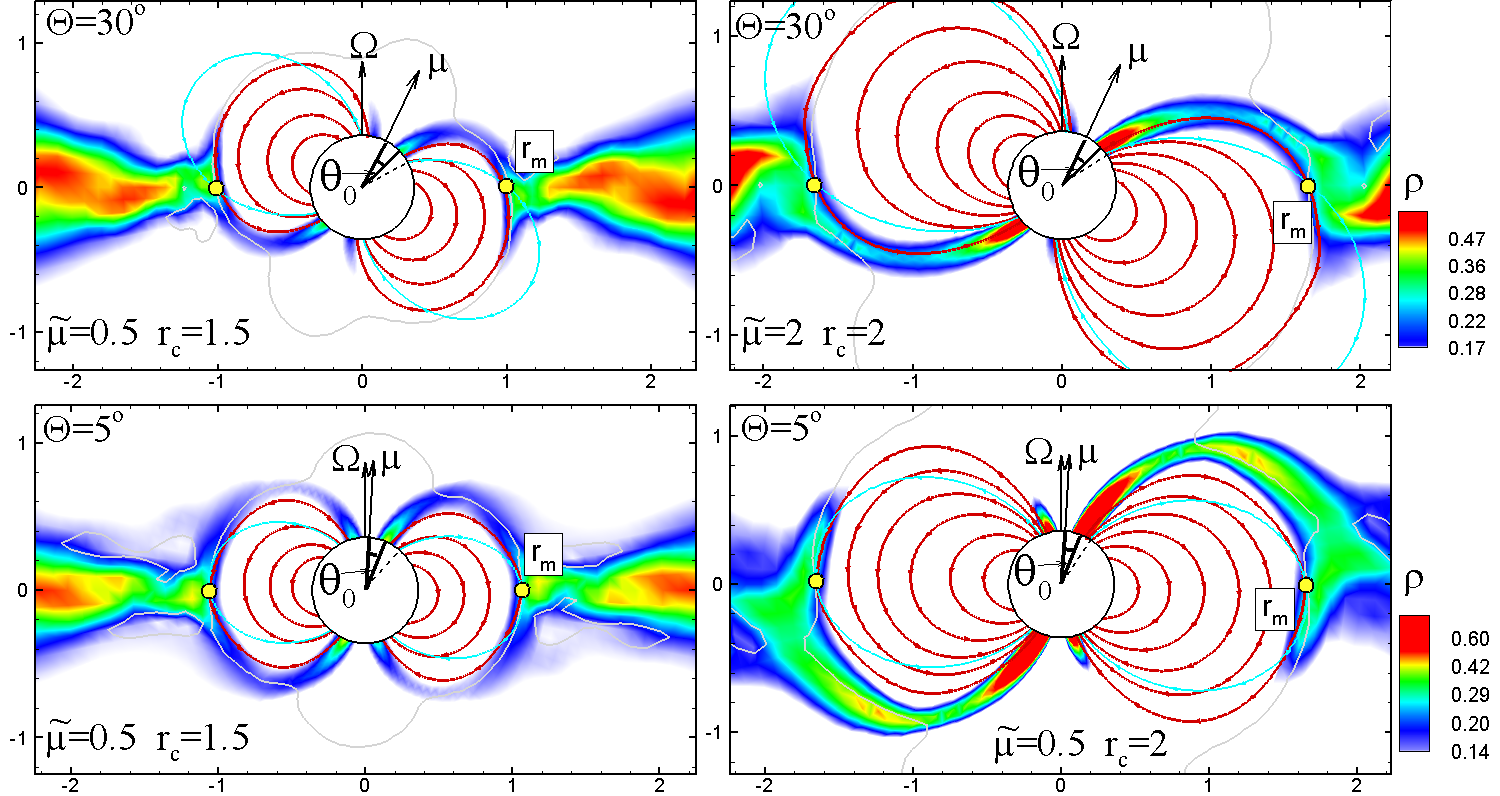}
\caption{The plots show that the polar angle $\theta_0$ (the angle between solid
lines) is approximately the same for different parameters of the star. Top and bottom
panels show cases of large ($\Theta=30^\circ$) and small ($\Theta=5^\circ$)
misalignment angles. The left and right panels show cases of small
($\widetilde{\mu}=0.5$) and large ($\widetilde{\mu}=1.8$) magnetic moments of the
star. Light-gray line shows The $\beta=1$ line is shown in light gray, and the yellow
dots show the position of the magnetospheric radius $r_m$. The dipole field lines,
which cross the magnetospheric radius are shown in cyan. The dashed black lines show
the theoretically-predicted polar position of the spot, $\theta_s$ (see Eq.
\ref{eq:theta_s-theoretical}).} \label{theta0-t5-t30-4}
\end{figure*}

\begin{figure*}
\centering
\includegraphics[width=8.5cm]{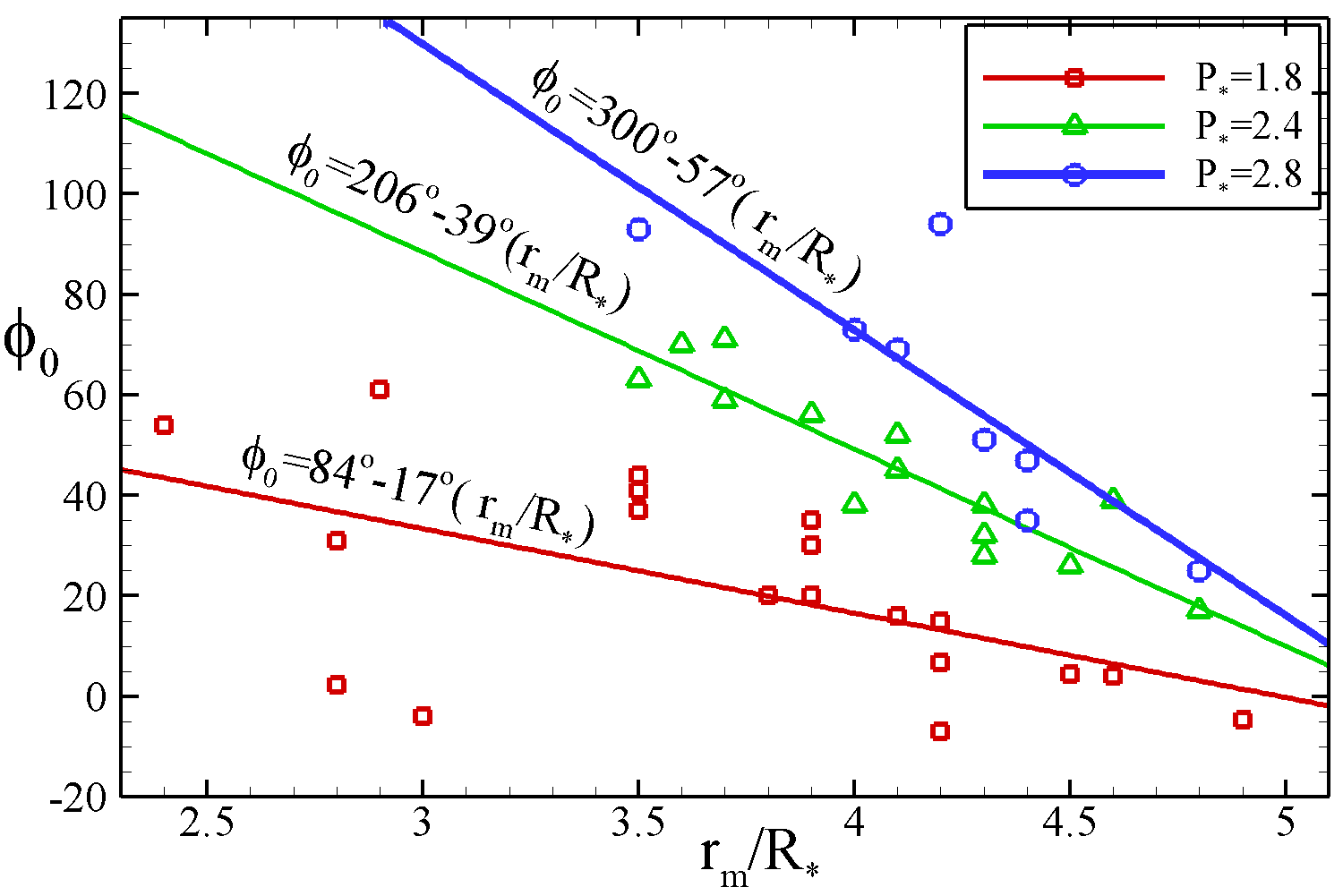}
\includegraphics[width=8.5cm]{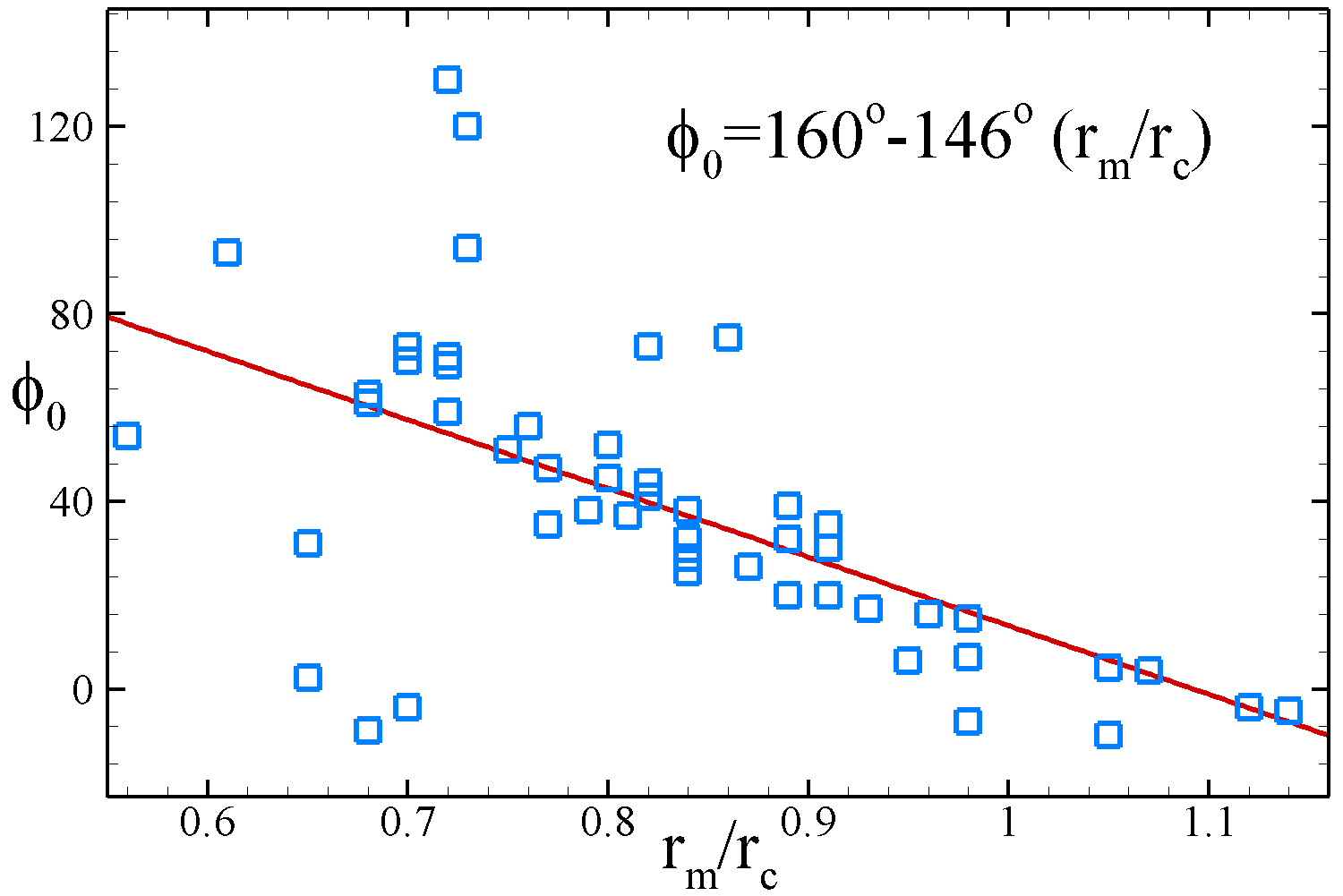}
\caption{\textit{Left Panel:} Dependence of $\phi_0$ on the magnetospheric radius
$r_m/R_\star$ for different periods of the star $P_\star$. The case of
$\Theta=2^\circ$ is removed from the set because the spot changes its position, and
its averaged preferred position has a large error. \textit{Right panel:} Dependence of
the azimuthal position of the spot $\phi_0$ on the ratio of magnetopsheric to
corotation radius $r_m/r_c$, taken for all simulation runs.} \label{phi0-rm-rc}
\end{figure*}

For each simulation run, we chose an interval of time where the accretion rate is
approximately constant. We derived an averaged spot using several moments in time from
this interval, then approximated the observed spot (energy flux distribution) with Eq.
\ref{eq:spotflux}, and derived best-fit parameters $(\theta_0, \phi_0)$, $(\theta_1,
\phi_1)$ and $\overline{F}_c$. Table \ref{tab:main} lists these parameters
 for different values of $\Theta$,
$r_m/R_\star$, and $P_\star$ ($r_c/R_\star$). Figures \ref{theta0-phi0},
\ref{theta1-phi1} and \ref{F0} show each of the five parameters of the spot for all
simulation runs (symbols) for different misalignment angles $\Theta$ (different
colors) and different periods of the star $P_\star$ (different symbols). Below, we
discuss each parameter and derive useful dependencies.

\subsection{Polar position of the spot, $\theta_0$}

We derive the polar position of the spot using two methods: (1) simple analytical
approach for a pure dipole field, and (2) numerical simulations.

\subsubsection{Polar position of the spot derived from the dipole model}

One of the parameters -- the polar position of the spot, $\theta_0$, -- can be derived
analytically.

Fig. \ref{dipole-spot} (left panel) shows a schematic view of a star with a dipole
field tilted at angle $\Theta$.
 The dipole field line which crosses the inner disc radius $r=r_m$ is shown.
This line also shows the expected position of the spot.
 The dipole field lines
obey the relation $r/\sin^2\theta =$ constant, where $\theta$ is measured from the
magnetic axis. For simplicity, let us assume that the center of the spot is in the
$\mu-\Omega$ plane.

Referring to Fig. \ref{dipole-spot}, we have
\begin{equation}
\cfrac{r_m}{\sin^2\left(\cfrac{\pi}{2}-\Theta\right)} =
\frac{R_\star}{\sin^2\theta_s},
\end{equation}
 which gives us the location of the spot, $\theta_s$, as
\begin{equation}
\theta_s = \sin^{-1} \left( \sqrt{\frac{R_\star}{r_m}} \cos\Theta \right).
\label{eq:theta_s-theoretical}
\end{equation}

In this idealized model, the derived angle $\theta_s$  shows the approximate position
of the spot in the polar direction, $\theta_0\approx\theta_s$.  The position of the
spot depends on truncation radius $r_m/R_\star$ and misalignment angle $\Theta$.

This could be a useful approach in finding the polar position of the spots. However,
numerical simulations show that the polar position is different from that derived
analytically, because the magnetic field is compressed by the disc.  Compression
changes the shape of the field lines,  affecting the position of the spots on the
star.


\subsubsection{Polar position of the spots derived from numerical simulations}

Table \ref{tab:main} and Fig. \ref{theta0-phi0} (left panel) show values of $\theta_0$
for all simulation runs. It can be seen that the polar position varies within a narrow
range of angles, $16^\circ\lesssim\theta_0\lesssim 18^\circ$. No any significant
correlation has been found between $\theta_0$ and the other parameters. This result is
somewhat unexpected, because according to the simple estimates performed in the
previous section, larger values of $\theta_0$ are expected for smaller misalignment
angles $\Theta$ and smaller magnetospheres (smaller $r_m/R_\star$), and vice-versa.
According to Eq. \ref{eq:theta_s-theoretical}, if we use extreme parameters of
$\Theta=2^\circ$ (the smallest  misalignment angle) and $r_m/R_\star=2.4$ (the
smallest magnetosphere), then we expect the largest angle to be $\theta_0=40.2^\circ$.
For  $\Theta=30^\circ$ and the largest magnetosphere, $r_m/R_\star=4.9$, we expect the
smallest angle to be $\theta_0=23^\circ$. Both angles are larger than the angles
obtained in our numerical simulations.

To understand this phenomenon, we compared the shapes of the magnetospheres obtained
in our simulations with the shape of the dipole magnetosphere. An example of such a
comparison is shown in Fig. \ref{dipole-spot} (right panel). Within the inner
magnetosphere, the field lines of the simulated magnetosphere (yellow lines) depart
only slightly from the dipole field line (shown in red). However, in the outer
magnetosphere, the field lines of the modeled magnetosphere strongly depart from the
dipole field lines. Matter flowing along the compressed field line forms a spot closer
to the magnetic pole (at smaller $\theta_0$), compared with the
theoretically-predicted position of the spot in case of a dipole field.

Fig. \ref{dipole-spot} also shows the position of the magnetospheric radius $r_m$,
where the $\beta=1$ line crosses the equatorial plane (see two black circles). It can
be seen that the field lines which start at $r_m$ connect the disc with the central
parts of the spots.

We also plot slices of density distribution in several cases with ``extreme"
parameters. Namely, we choose those simulation runs where  parameters $\Theta$ and
$r_m/R_\star$ are very large or very small. Thus, according to Eq.
\ref{eq:theta_s-theoretical}, the spots are expected to be at different polar
positions $\theta_s$. Fig. \ref{theta0-t5-t30-4} shows  that in spite of the
difference in parameters,  the polar position of the spot $\theta_0$ is approximately
the same and corresponds to the narrow range of values observed in Fig.
\ref{theta0-phi0}. This can be explained by the fact that in cases of a large
magnetosphere compression is not very strong, but is sufficient enough to shift the
spot position towards the magnetic pole and yield a value of  $\theta_0$ which is
different from that predicted theoretically in the case of a dipole field. In cases of
small magnetospheres, theoretically-predicted angle $\theta_s$ is larger than in cases
of a large magnetosphere. However, compression of the magnetosphere is stronger, and
the spot has a stronger shift towards the magnetic pole, compared with cases of a
large magnetosphere. This effect leads to similar angles $\theta_0$ in our simulation
set.
It is clear that at even smaller (than in our set)  magnetospheres, the theoretical
dependence will not work, due to an even stronger compression of the magnetosphere.
However, the formula may be applicable
 at very large $r_m/R_\star$, where compression is expected to be weaker.

\subsection{Azimuthal position of the spot, $\phi_0$}

Figure \ref{theta0-phi0} (right panel) shows the azimuthal position of the spots for
all simulation runs. The figure shows a wide range of angles $\phi_0$. We noticed that
some correlation can be seen between groups of runs with the same period of the star
$P_\star$ (which are marked with the same symbol in the figure). To better demonstrate
this apparent correlation, we divided all data into three groups corresponding to
three periods of the star, and obtained much clearer dependencies of $\phi_0$ on
$r_m/R_\star$.  The list-square approximation shows linear correlations:

\begin{eqnarray}
&& \phi_0=84^\circ-17^\circ \frac{r_m}{R_\star}, ~~~ for ~~~ P_\star=1.8 ~~\left(\frac{r_c}{R_\star}=4.3\right)\\
&& \phi_0=206^\circ-39^\circ \frac{r_m}{R_\star}, ~~~ for ~~~ P_\star=2.4~~\left(\frac{r_c}{R_\star}=5.1\right)\\
&& \phi_0=300^\circ-57^\circ \frac{r_m}{R_\star}, ~~~ for ~~~
P_\star=2.8~~\left(\frac{r_c}{R_\star}=5.7\right)
\end{eqnarray}

It can be seen that angle $\phi_0$ varies more rapidly in the cases of a slower-
rotating star ($P_\star=2.8$), and more slowly in cases of a more rapidly-rotating
star ($P_\star=1.8$).

These dependencies can be interpreted as an azimuthal phase-shift that varies with
accretion rate: smaller values of $r_m/R_\star$ correspond to higher accretion rates
and stronger shifts $\phi_0$.

The azimuthal position of the spot depends on the relative velocity of the inner disc
with respect to the magnetosphere. Hence, the dependence of $\phi_0$ on the ratio
$r_m/r_c$ is expected.  In this case, we were able to obtain a unique correlation for
all simulation runs. Fig. \ref{phi0-rm-rc} (right panel) shows this correlation. The
least-square fit shows a linear dependence:

\begin{equation}
\phi_0=160^\circ-146^\circ \frac{r_m}{r_c}. \label{eq:Phi0-rmrc-line}
\end{equation}

Therefore, if a value of $r_m/r_c$ is known, then the approximate azimuthal position
of the spot is known as well.

\begin{figure*}
\centering
\includegraphics[width=16cm]{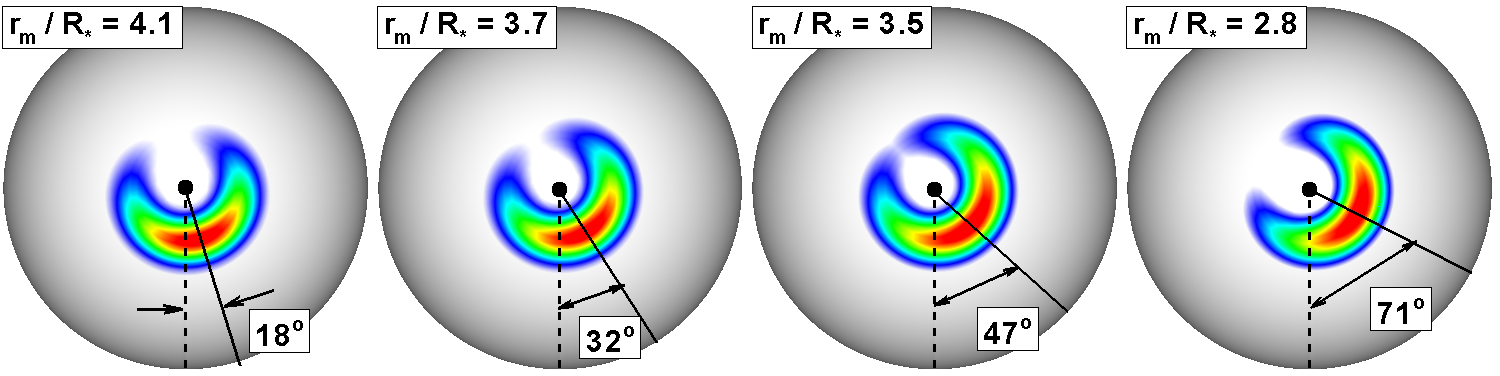}
\caption{The plot demonstrates different azimuthal positions $\phi_0$ of the spots at
different magnetospheric radii $r_m/R_\star$. The spot shifts in the azimuthal
direction   when $r_m/R_\star$ decreases. The colored background shows the energy
flux, which varies from the minimum value $F_{\rm min}=0.5$ (dark cyan color in all
panels) to the maximum value $F_{\rm max}$ (red color), which is 5.8, 4.8, 3.2 and
1.9, from left to right.} \label{spots-phase-4}
\end{figure*}

Fig. \ref{spots-phase-4} demonstrates how the spot position $\phi_0$ varies with
 $r_m/R_\star$ (which can be interpreted as variation of the accretion
rate): angle $\phi_0$ increases when $r_m/R_\star$ decreases.

We should note that in the case of very small misalignment angle, $\Theta= 2^\circ$,
the spot changes its position rapidly, making entire  $2\pi$ cycle around the magnetic
pole. Hence, the azimuthal position of the spot has a large error. This is why we
excluded the case of $\Theta=2^\circ$ from the left panel of Fig. \ref{phi0-rm-rc}.

\subsection{Polar width of the spot, $\theta_1$}

Figure \ref{theta1-phi1} (left panel) shows that the polar width of the spots varies
within a relatively narrow interval, $7^\circ\lesssim\theta_1\lesssim 10^\circ$, and
is   $\theta_1=7^\circ-8^\circ$ for most models. In a few cases, it is larger:
$\theta_1=9^\circ$, and $\theta=10^\circ$ in one simulation run. There is a weak
correlation between $\theta_1$ and $r_m/R_\star$, with $\theta_1$ being smaller in the
cases of larger $r_m/R_\star$.

\subsection{Azimuthal width of the spot, $\phi_1$}

Figure \ref{theta1-phi1} (right panel) shows the azimuthal width of the spots $\phi_1$
for all simulation runs.  This parameter is also scattered significantly, like
$\phi_0$. However, some ordered variation can be seen if we consider the subsets of
runs with the same misalignment angle $\Theta$ (marked by different colors). The
points with a large $\Theta$ are located at the bottom part of the plot, where
$\phi_1\approx 100^\circ$ and the spots have the shape of an arc. The points with
smallest $\Theta$ are located towards larger $\phi_1$, and have the shape of a ring.

\begin{figure*}
\centering
\includegraphics[width=8.5cm]{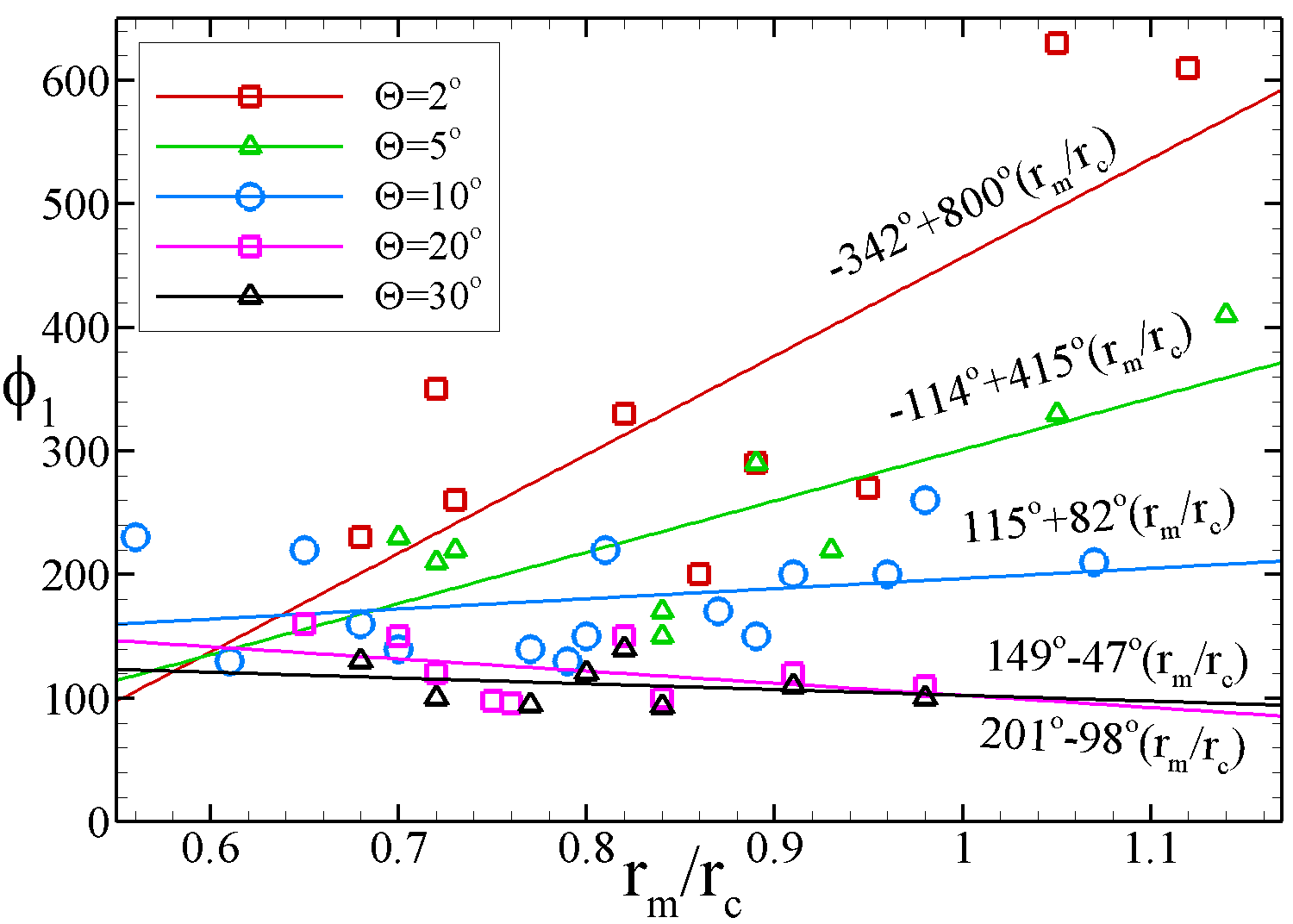}
\includegraphics[width=8.5cm]{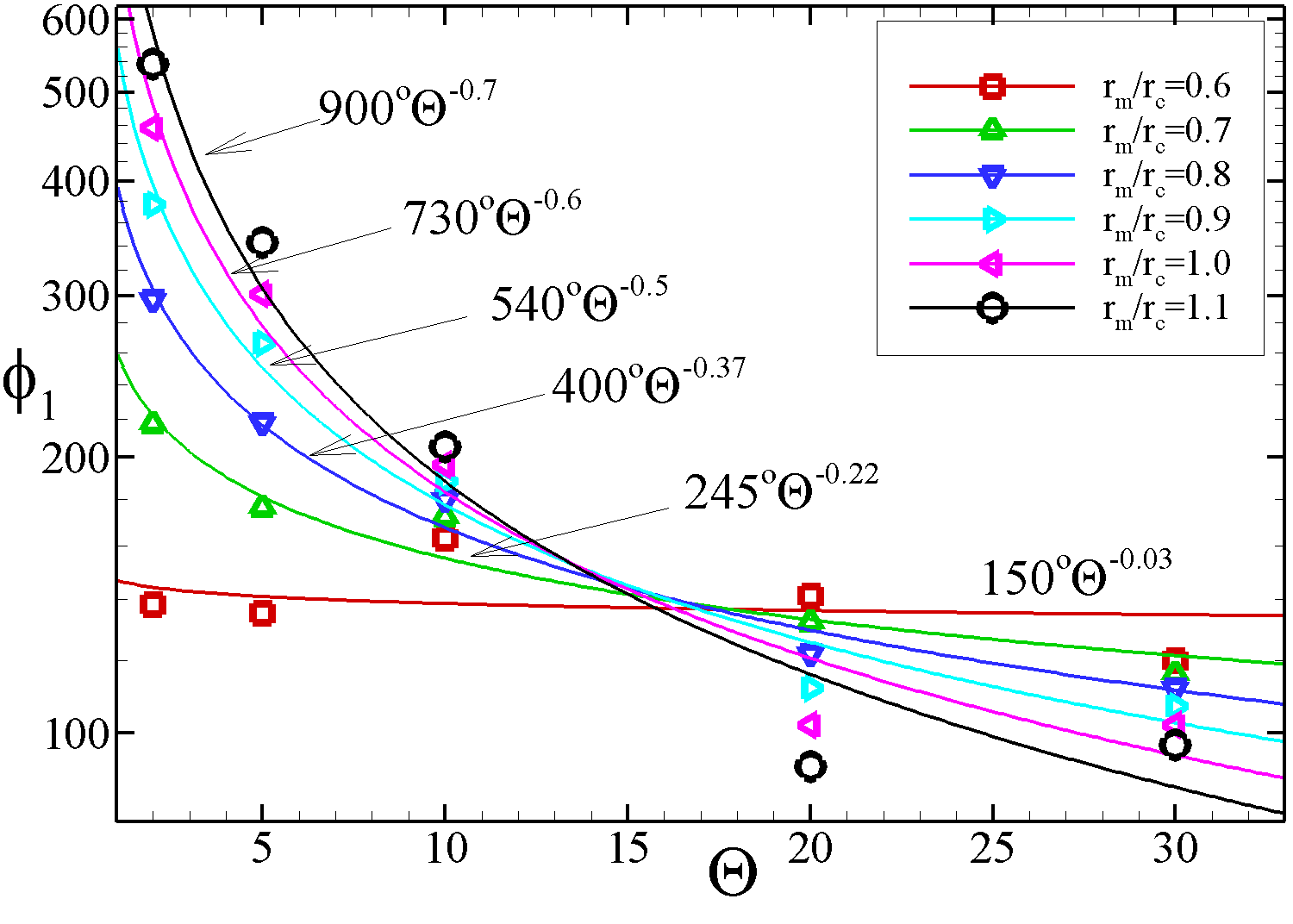}
\caption{\textit{Left Panel:}  Dependence of the azimuthal width of the spot $\phi_1$
on the ratio of magnetic to corotation radius $r_m/r_c$ for different misalignment
angles $\Theta$. \textit{Right Panel:} Dependence of $\phi_1$ on the misalignment
angle of the dipole moment $\Theta$ for different $r_m/r_c$.} \label{phi1-rm-rc}
\end{figure*}

Taking into account this (expected) dependence of $\phi_1$ on $\Theta$, we separate
all the runs into subsets with a different $\Theta$. We notice that the correlations
become stronger when we plot $\phi_1$  as a function of $r_m/r_c$. Fig.
\ref{phi1-rm-rc} (left panel) shows subsets of data for each $\Theta$ and a linear
approximation for each subset. At large misalignment angles,
$\Theta=20^\circ-30^\circ$, the azimuthal width does not vary much with $r_m/r_c$, and
$\phi_1\approx 90^\circ-100^\circ$. However, $\phi_1$ strongly increases with
$r_m/r_c$ for small $\Theta$. The lines which approximate sets of data have different
slopes, and thus, the dependencies on $\Theta$ are different. To derive these
dependencies, we choose several values of $r_m/r_c$ and take the values of $\phi_1$
that correspond to lines at different $\Theta$. We plot these values in Fig.
\ref{phi1-rm-rc} (right panel) and also obtain the following dependencies:

\begin{eqnarray}
&& \phi_1=150^\circ \Theta^{-0.03}, ~~~ for ~~~ r_m/r_c=0.6~, \\
&& \phi_1=400^\circ \Theta^{-0.37}, ~~~ for ~~~ r_m/r_c=0.8~, \\
&& \phi_1=730^\circ \Theta^{-0.60}, ~~~ for ~~~ r_m/r_c=1.0~.
\label{eq:phi1-Theta}
\end{eqnarray}

The Figure and Eqs. 8-10 show that for  $\Theta\gtrsim 15^\circ$, the width of the
spots is relatively small for all values of $r_m/r_c$. However, for $\Theta\lesssim
15^\circ$, the shape varies from
 an arc for smaller $r_m/r_c$ to a
ring for larger $r_m/r_c$. For a star with the same corotation radius $r_c$,
larger/smaller values of $r_m/r_c$ correspond to a smaller/larger accretion rate.

\begin{figure*}
\centering
\includegraphics[width=16.0cm]{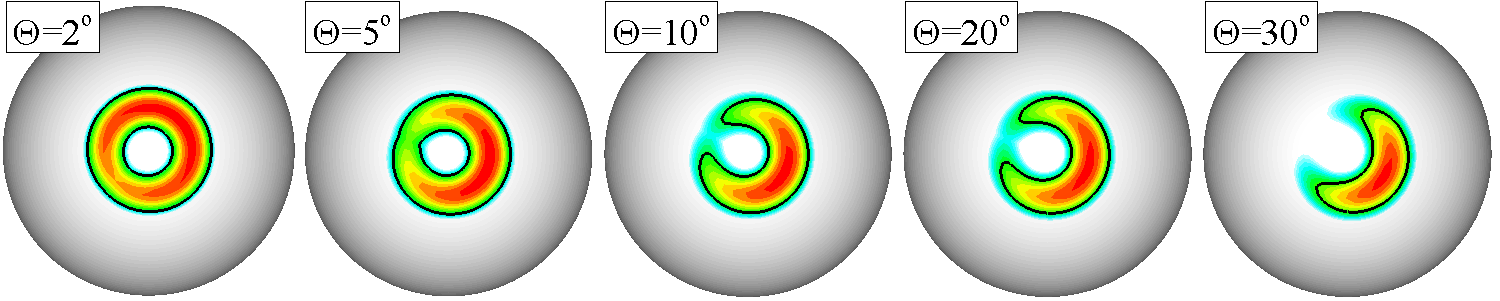}
\caption{Shapes of the spots in cases of different $\Theta$. The colored background
shows the energy flux distribution. The black lines indicate the boundary where the
energy flux is $e$ times smaller than the maximum flux. Other parameters are:
$\widetilde{\mu}=1$, $r_c=1.8$, and are the same in all cases. The energy levels vary
from the minimum value of $F_{\rm min}=0.5$ (light-cyan color), which is the same in
all cases, to the maximum value of $F_{\rm max}$ (red color), which is $2.1$ in the
three left panels, and is $3.0$ and $3.8$ in the 4th and 5th panels, respectively.}
\label{spots-Theta-5}
\end{figure*}

Fig. \ref{spots-Theta-5} demonstrates the shape of the spots in the case of the same
magnetic moment $\widetilde{\mu}$ and period $P_\star$ of the star (approximately the
same value of $r_m/r_c$), but a different misalignment angle of the dipole $\Theta$.
As we can see, the shapes of the spots vary from a ring (at small $\Theta$) to an arc
(at large $\Theta$).

\subsection{Energy flux in the center of the spots, $F_c$}

From our simulations and fits of the observed spots with Eq. \ref{eq:spotflux}, we
obtain the maximum energy flux in the center of the spot $\widetilde{F}_c$ (in
dimensionless units). The dimensional flux $F_c$, which is used in Eq.
\ref{eq:spotflux}, can be derived from Eq. \ref{eq:flux}. It can be seen that the main
dimensionless variable in Eq. \ref{eq:flux} is the ratio
$\overline{F}_c=\widetilde{F_c}/\widetilde\mu^2$. This is why we put this variable
into the Table \ref{tab:main} and use it in Fig. \ref{F0}.

Fig. \ref{F0} shows that the values of $\overline{F}_c$ are mainly confined to the
interval $0.5\lesssim \overline{F}_c\lesssim 4$. It is larger, $\overline{F}_c\approx
6.5-8.5$, in a few points corresponding to a smaller magnetosphere,
$r_m/R_\star\approx 2.8-3.2$. The fluxes are about two times smaller in cases of a
small $\Theta$ (see red line in Fig. \ref{F0}).

\begin{table*}
\begin{tabular}{l@{\extracolsep{0.2em}}l@{}lllll}
\hline
$P_\star$ (dim) ~~~ &  $r_c$ ~~~   & $r_c/R_\star$    & $P_\star$ (CTTSs)  & $P_\star$ (White dwarfs)    & $P_\star$ (Neutron stars)   \\
\hline
1.8          ~~~&   1.5     ~~~ &    4.3    &          3.2 days       & 52 s             & 4.0 ms                 \\
2.4         ~~~ &   1.8     ~~~ &    5.1    &          4.3 days       & 70 s             & 5.3 ms                 \\
2.8         ~~~ &   2.0    ~~~  &    5.7    &          5.0 days       & 81 s             & 6.2 ms                 \\
\hline
\end{tabular}
\caption{From left to right: dimensionless period of the star $P_\star$, dimensionless
corotation radius $r_c$; corotation radius normalized to stellar radius,
$r_c/R_\star$; dimensional period of the star $P_\star$, for different types of
stars.} \label{tab:P-rc}
\end{table*}

\section{Magnetospheric radius}
\label{sec:magnetospheric_radius}

Magnetospheric radius  $r_m$ (which corresponds to the boundary between the
magnetosphere and the disc) is often derived from the condition that it is
proportional to the Alfv\'en radius $r_A^0$, derived for spherical accretion under the
assumption that the magnetic pressure in the magnetosphere equals to the ram pressure
of
  free-falling matter (e.g., \citealt{LambPethickPines73}):
\begin{equation}
r_m = k_A r_A^0~, ~~~~~~ r_A^0=\left( \frac{\mu^4}{2GM_\star\dot{M}^2} \right)^{1/7}~,
\label{eq:Alfven}
\end{equation}
where $k_A$ is the dimensionless coefficient of the order of unity.  This formula has
been tested a few times. In these tests, the inner disc radius has been derived in two
different ways. First, it has been derived from the simulations using conditions
$\beta=1$ or $\beta_1=1$. Second, it has been derived using formula \ref{eq:Alfven}
for $r_A^0$, where values such as $\mu$ and $\dot M$ were taken from the simulations.
\citet{LongRomanovaLovelace05} performed such comparisons in three models with
slightly different parameters $\mu$ and the condition $\beta_1=1$, and obtained an
average approximate value of
 $k_A\approx 0.5$. \citet{BessolazEtAl08} used a variety of definitions for the inner
disc radius, such as $\beta=1$ and $\beta_1=1$, and obtained  $k_A\approx 0.4$ in the
case of the $\beta=1$ condition (see also \citealt{ZanniFerreira13}).
 However, none of this work studied the dependence of $r_m$ on $\mu$ or $\dot M$.

We have a sufficiently large set of simulations, so these dependencies can be tested
here. As the first step, we take the formula for the Alfv\'en radius (Eq.
\ref{eq:Alfven}), substitute in our dimensional units (see Sec. \ref{app:refval}), and
derive the magnetospheric radius in a dimensionless form:
\begin{equation}
\frac{r_m}{R_\star}=\frac{k_A}{0.35} \bigg(\frac{\widetilde{\mu}^2}{\widetilde{\dot
M}}\bigg)^{2/7}~ . \label{eq:rm-theoretical}
\end{equation}

\begin{figure}
\centering
\includegraphics[width=8.5cm]{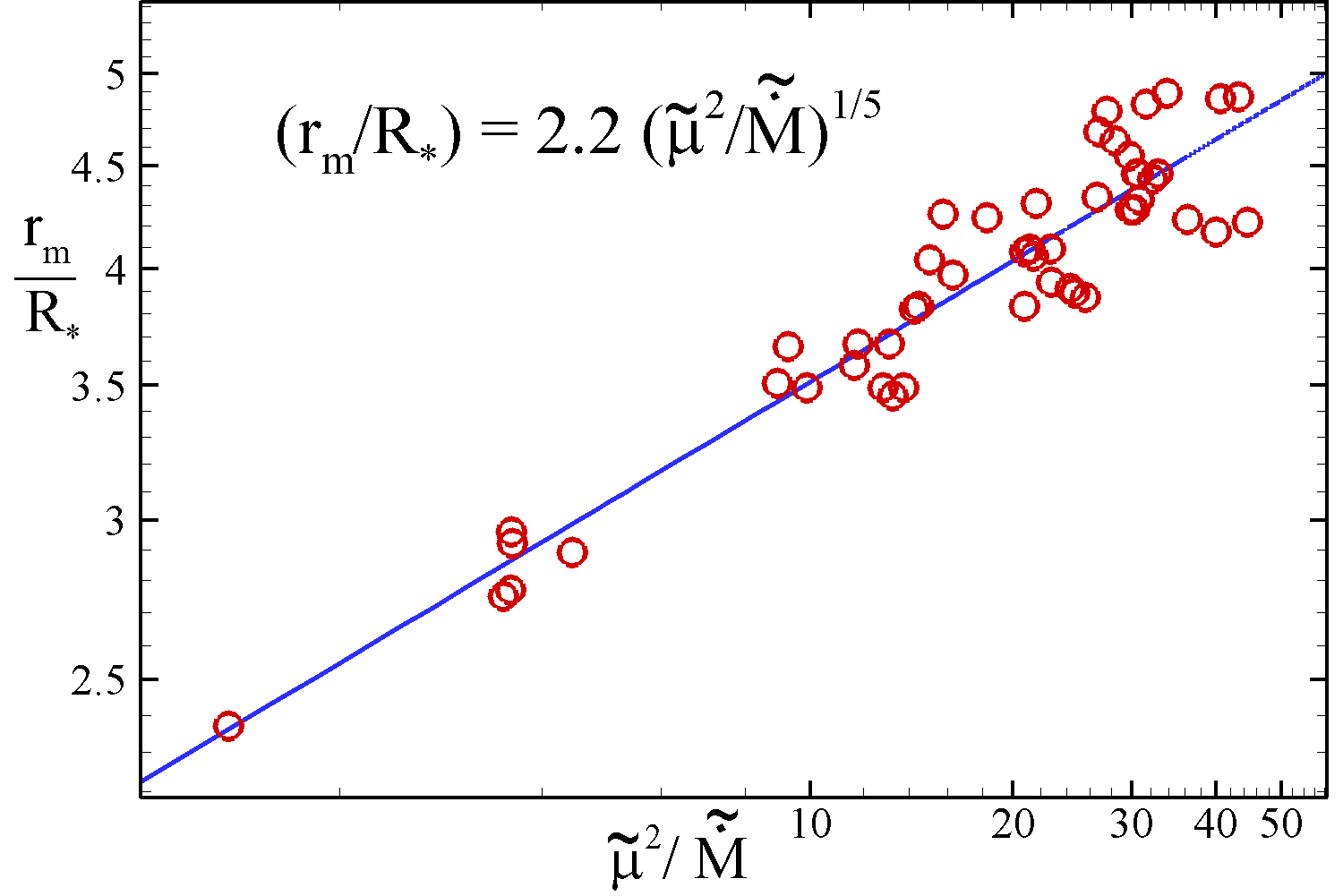}
\caption{Normalized magnetospheric radius $r_m/R_\star$ as a function of the ratio
$\widetilde{\mu}^2/{\widetilde{\dot M}}$ for all simulation runs (circles). The line
shows the power fit approximation.} \label{rm-mu2mdot}
\end{figure}

 We can test this formula in our simulations, where the dimensionless magnetic moment
$\widetilde{\mu}$ is the main initial parameter in each simulation run, and the
dimensionless accretion rate $\widetilde{\dot M}$ is the output from each simulation
run. We also derived the ratio $r_m/R_\star$ from the condition that $\beta=1$ in each
simulation run.  Fig. \ref{rm-mu2mdot} shows the resulting dependence of $r_m/R_\star$
on ${\widetilde{\mu}^2}/{\widetilde{\dot M}}$. The dependence is well-approximated by
the power law in the following form:

\begin{equation}
\frac{r_m}{R_\star}\approx 2.2 \bigg(\frac{\widetilde{\mu}^2}{\widetilde{\dot
M}}\bigg)^{1/5}~. \label{eq:rm-derived}
\end{equation}

We can see that the power obtained from the simulations, $a\approx 0.2=1/5$ is smaller
than the power given by the formula for the Alfv\'en radius, $a=2/7\approx 0.29$. We
suggest that the compression of the magnetosphere is probably responsible for
different dependence of $r_m$ on $\widetilde{\mu}^2/\widetilde{\dot M}$. Namely, when
the magnetosphere is compressed, the magnetospheric radius varies slower with magnetic
moment and accretion rate.

From the comparison of equations \ref{eq:rm-derived} and \ref{eq:rm-theoretical} we
can derive coefficient $k_A$, which is not a constant anymore:

\begin{equation}
k_A\approx 0.77 \bigg(\frac{\widetilde{\mu}^2}{\widetilde{\dot M}}\bigg)^{-0.086}.
\end{equation}

The dependence of $k_A$ on $\widetilde{\mu}^2/\widetilde{\dot M}$ is  weak, so that
for the range of $2\lesssim\widetilde{\mu}^2/\widetilde{\dot M}\lesssim 50$ used in
our set, we obtain $k_A$ in the range of $0.55\lesssim k_A\lesssim 0.72$. These values
of parameter  $k_A$ are in agreement with the values derived earlier in axisymmetric
simulations (e.g., \citealt{LongRomanovaLovelace05}).

Next, we use Eq. \ref{eq:rm-derived} and Eq. \ref{eq:mdot} to derive the ratio
$r_m/R_\star$ through dimensional parameters:

\begin{equation}
\frac{r_m}{R_\star} \approx 1.06 \left( \frac{\mu^4}{\dot{M}^2 GM_\star
R_\star^7}\right)^{1/10} ~.  \label{eq:rm-derived-dimensional}
\end{equation}

\noindent  This formula can be used if the magnetic moment of the star and the other
parameters are known.

We should note that the  dependence derived in Eq. \ref{eq:rm-derived} is relevant for
the magnetospheric radii $2.5\lesssim r_m/R_\star\lesssim 5$ investigated in this
work.
 A separate analysis should be done for larger and smaller
radii. We anticipate that at larger $r_m/R_\star$ compression will be less important
and the power in Eq. \ref{eq:rm-derived-dimensional} will be larger than $0.2$, while
at smaller $r_m/R_\star$ compression will be more important and the power will be
smaller.
 Special simulation runs should be performed to
investigate this dependence at larger and smaller $r_m/R_\star$.

\section{Some Practical Considerations}

The above analysis shows that the shapes and positions of the spots are determined by
three parameters: (1) the misalignment angle of the dipole moment $\Theta$, (2) the
normalized corotation radius $r_c/R_\star$ and (3) the normalized magnetospheric
radius $r_m/R_\star$. Observations of magnetized stars are usually complex, and most
of the parameters of the star and the inner disc are not known, or can be estimated
only approximately. Below, we discuss possible ways to derive these parameters.

\begin{itemize}

\item It is often the case that the period of the star $P_\star$ is one of the
best-known parameters. If mass $M_\star$ and radius $R_\star$ of the star are known,
then the dimensionless period $P_\star$ and the ratio $r_c/R_\star$ can be derived.

\item Many evolved magnetized stars are expected to be in the rotational equilibrium
state, where a star's spin varies around the value corresponding to rotational
equilibrium. In these cases, the magnetospheric radius can be derived from condition
$r_m=k r_c$ (where $k\approx 1$) \footnote{The condition $k=1$ corresponds to the case
where the magnetosphere interacts only with the innermost radius of the disc at
$r=r_m$. In reality, stellar angular momentum may partially flow to the corona along
inflated field lines, and partially to the disc along the field lines connecting a
star with the disk at $r>r_m$. Therefore, it is expected that $k<1$ in rotational
equilibrium state.}. If a star is not in rotational equilibrium, then this condition
cannot be used.

In another approach, if the magnetic moment of the star and the accretion rate are
known, then $r_m$ can be calculated  from the formula for the Alfv\'en radius derived
for spherical accretion (see Eq. \ref{eq:Alfven}). In this paper, we perform an
independent derivation of a similar formula and find that it differs from the standard
formula for the Alfv\'en radius (see Sec. \ref{sec:magnetospheric_radius}).

\item One of the most important parameters, $\phi_0$, which stands for the azimuthal
shift of the spot, depends only on the ratio $r_m/r_c$ (see Eq. \ref{phi0-rm-rc}). If
a star is in the rotational equilibrium state, then the value $r_m/r_c\approx$ can be
taken from numerical simulations. Axisymmetric simulations performed by
\citet{LongRomanovaLovelace05} show that $r_m/r_c\approx 0.7-0.8$, while
\citet{ZanniFerreira13} found $r_m/r_c\approx 0.6$. This ratio is smaller, when a
larger amount of angular momentum flows into the corona along inflated or
partially-inflated field lines. This ratio can also vary in time. Therefore, the above
numbers are approximate and can be used only as estimates.

\item The misalignment angle of the dipole's magnetic moment $\Theta$, is known
approximately in some cases. For example, in a few CTTSs,  the magnetic field
distribution on the surface of the star has been derived from polarimetric
observations and then approximated with the set of tilted magnetic moments of a
different order. It was found that the misalignment angle of the dipole component is
usually small, $\Theta\approx 10^\circ-20^\circ$ (e.g., \citep{DonatiEtAl08}). In some
cases, the misalignment angle can be obtained from the shapes of observed light-curves
and comparisons with those obtained in 3D MHD numerical simulations
\citep{RomanovaEtAl04}. For example, for some probable inclination angle of $i\approx
45^\circ$, a nearly-symmetric light-curves with one peak per period yield a relatively
small misalignment angles, $\Theta\lesssim 30^\circ$, while light-curves with two
peaks per period yield a larger $\Theta$.

\end{itemize}

\section{Conclusions and Discussion}

In this work we derived a useful analytical formula (Eq.  \ref{eq:spotflux}) that
provides an approximation for arc-shaped and ring-shaped spots which form in accreting
magnetized stars. Multiple 3D MHD simulation runs were performed to find the positions
of the spots ($\theta_0$, $\phi_0$), their shapes ($\theta_1$, $\phi_1$) and their
dependencies on the period of the star $P_\star$, the misalignment angle of the dipole
$\Theta$, and the ratio of the magnetospheric radius to the radius of the star,
$r_m/R_\star$. The main conclusions are as follows:

\textbf{1.} The polar positions of the spots $\theta_0$ vary within the narrow range
of $16^\circ\lesssim\theta_0\lesssim 18^\circ$. These angles are different from the
polar angles obtained with the simple analytical formula (see Eq.
\ref{eq:theta_s-theoretical}), where $\theta_0$ has larger values and varies
considerably with $r_m/R_\star$ and the dipole misalignment angle $\Theta$. We noticed
that in all our numerical models the external closed magnetosphere is compressed and
 departs significantly from the dipole shape, which leads to the misplacement of the spot.

\textbf{2.} The azimuthal positions of the spots $\phi_0$ strongly vary with the
magnetospheric radius $r_m/R_\star$ (accretion rate). They increase, when $r_m$
decreases. For the same interval of $r_m/R_\star$, $\phi_0$ varies more rapidly for
stars with larger periods.

\textbf{3.} An important correlation has been found for the dependence of $\phi_0$ on
the ratio $r_m/r_c$, which is valid for all simulation runs.

\textbf{4.} The polar widths of the spots $\theta_1$ vary within the narrow interval
of $7^\circ\lesssim\theta_1\lesssim 9^\circ$, and  the majority of the spots have a
width of $7^\circ-8^\circ$.

\textbf{5.} The azimuthal widths of the spots $\phi_1$ strongly depend on the dipole
misalignment angle $\Theta$. They vary from $\phi_1=90^\circ-100^\circ$ for
$\Theta=30^\circ$ (arc-shaped spots) to $\phi_1\approx 600^\circ$ for $\Theta=2^\circ$
(ring-shaped spots).

\textbf{6.}
 We also used our data to check the formula for the Alfv\'en
radius, where the main dependence is $r_m\sim (\mu^2/{\dot M})^{2/7}$. We found that
the dependence is more gradual, $r_m\sim (\mu^2/{\dot M})^{1/5}$, which can be
explained by the compression of the magnetosphere by the disc matter and by the
non-dipole shape of the magnetic field lines of the external magnetosphere.

\smallskip

Once these parameters are known, the analytical formula given by Eq. \ref{eq:spotflux}
can be used to model the energy flux distribution in the spot. The light-curves can be
calculated using a separate numerical program. In this program, the radiation model,
the  anisotropy of the radiation, and the inclination angle of the system can be taken
into account.   For neutron stars, this code should take into account relativistic
light-bending, Doppler-shift and other relativistic effects (see, e.g.,
\citealt{KulkarniRomanova05}).

Variation of  the azimuthal position of the spot $\phi_0$ with accretion rate is
particularly important for observations.  For example, in millisecond pulsars, it may
lead to phase-shifts in light-curves \citep{PatrunoEtAl09} and timing noise (e.g.,
\citealt{PoutanenEtAl09}). In CTTSs, it can lead to the phenomenon of a drifting
period, where the period derived at different times of observation, is different, and
therefore, it varies with time (e.g., \citealt{RucinskiEtAl08}).

Here, we present results for a relatively narrow interval of the ratio $0.6\lesssim
r_m/r_c\lesssim 1.1$, which corresponds to ordered accretion through two funnel
streams. The smallest values correspond to a boundary between stable and unstable
regimes of accretion \footnote{Recent simulations on a grid twice as fine show that
this boundary occurs at somewhat larger values of $r_m/r_c$ (Romanova et al. 2013, in
prep.)}, where at $r_m/r_c\lesssim 0.6$ accretion proceeds in the equatorial plane
through the magnetic Rayleigh-Taylor instability. The largest value, $r_m/r_c\approx
1.1$, corresponds to the propeller regime, in which the funnel accretion is forbidden
by the centrifugal barrier (e.g., \citealt{IllarionovSunyaev75, LovelaceEtAl99}).

Of course, the analytical formula (Eq. \ref{eq:spotflux}) describing the shape of the
spot can be applied only in the cases where the magnetic field of the star has a
strong dipolar component. In the case of a more complex field, the higher order
component of the field may govern the funnel flow near the star, and the shapes of the
spots will be determined by higher order multipoles of the field (e.g.,
\citealt{LongEtAl11, RomanovaEtAl11}). This case should be investigated separately.


\subsection*{Acknowledgments}
{Authors thank Alisa Blinova for help and Alexander Koldoba for an earlier-developed
`cubed sphere' code. Resources supporting this work were provided by the NASA High-End
Computing (HEC) Program through the NASA Advanced Supercomputing (NAS) Division at
Ames Research Center and the NASA Center for Computational Sciences (NCCS) at Goddard
Space Flight Center.  The research was supported by NASA grants NNX10AF63G, NNX11AF33G
and NSF grant AST-1008636.}


\bibliography{ms}

\appendix

\section{Reference Values}
\label{app:refval}

\begin{table}
\begin{tabular}{l@{\extracolsep{0.2em}}l@{}lll}

\hline
&                                                   & CTTSs       & White dwarfs          & Neutron stars           \\
\hline

\multicolumn{2}{l}{$M_\star(M_\odot)$}              & 0.8         & 1                     & 1.4                     \\
\multicolumn{2}{l}{$R_\star$}                       & $2R_\odot$  & 5000 km               & 10 km                   \\
\multicolumn{2}{l}{$B_\star$ (G)}                   & $10^3$      & $10^6$                & $10^9$                  \\
\multicolumn{2}{l}{$\mu$}                           & 1           & 1                     & 1                       \\
\multicolumn{2}{l}{$R_0$ (cm)}                      & $4\e{11}$   & $1.4\e9$              & $2.9\e6$                \\
\multicolumn{2}{l}{$P_0$}                            & 1.8 days  & 29 s                  & 2.2 ms                \\
\multicolumn{2}{l}{$B_0$ (G)}                       & 43          & $4.3\e4$              & $4.3\e7$                \\
\multicolumn{2}{l} {$\dot M_0$ ($M_\odot$yr$^{-1}$)}   & $2.8\e{-7}$   & $1.9\e{-7}$       & $2.9\e{-8}$             \\
\multicolumn{2}{l}{$F_0$ (erg s$^{-1}$cm$^{-2}$)}   & $3\e{10}$ & $5.6\e{17}$             & $1.5\e{25}$             \\
\hline
\end{tabular}
\caption{Sample reference values of the dynamical quantities used in our simulations
in the case where dimensionless magnetic moment $\widetilde{\mu}=1$. Note that
parameters $\rho_0$, $B_0$, $\dot M_0$ and $F_0$ depend on $\widetilde{\mu}$ (see Sec.
\ref{app:refval}).} \label{tab:refval}
\end{table}

Our simulations are done using dimensionless variables (denoted here by a tilde),
obtained by dividing the dimensional variables by  reference values. The reference
values are determined as follows: We choose fiducial values for stellar mass
$M_\star$, radius $R_\star$ and equatorial surface magnetic field $B_\star$; the
reference values of all other dynamical quantities are then obtained from these
fiducial values. The unit of distance is chosen to be $R_0 = R_\star/0.35$. The
reference velocity is the Keplerian velocity at $R_0$, $v_0 = (GM_\star/R_0)^{1/2}$,
and $\omega_0 = v_0/R_0$ is the reference angular velocity. The reference time is the
Keplerian rotation period at $R_0$, $P_0 = 2\pi R_0/v_0$.

The reference magnetic moment $\mu_0$ = $\mu/\widetilde{\mu}$, where $\widetilde{\mu}$
is the dimensionless magnetic moment of the star. The reference magnetic field is $B_0
= \mu_0/R_0^3$. The reference density is $\rho_0 = B_0^2/v_0^2$. The reference
accretion rate is $\dot{M}_0 = \rho_0 v_0 R_0^2$. Using the relationships for
$\rho_0$, $B_0$ and $v_0$, we obtain $\dot{M}_0 = \mu^2 / (\widetilde{\mu}^2 v_0
R_0^4)$. The dimensional accretion rate is $\dot{M} = \widetilde{\dot{M}}\dot{M}_0$,
where $\widetilde{\dot M}$ is the dimensionless accretion rate onto the surface of the
star which is obtained from the simulations. Substituting the expression for
$\dot{M}_0$ and using $\mu = B_\star R_\star^3$, we get
\begin{equation}
\label{eq:mdot} \dot{M} = \left( \frac{ \widetilde{\dot{M}} }{ \widetilde{\mu}^2 }
\right) \frac{B_\star^2 R_\star^{5/2} (0.35)^{7/2}}{  (GM_\star)^{1/2} }.
\end{equation}
Thus, if the stellar parameters $M_\star$, $R_\star$ and $B_\star$ are fixed, then the
accretion rate is determined by $\widetilde{\dot{M}}/\widetilde{\mu}^2$. Therefore,
variation of parameter $\widetilde\mu$ can be interpreted as variation of the
magnetospheric radius of the star, or it can be equally validly interpreted as a
parameter that is responsible for variation of the accretion rate.

The reference energy flux is defined as $F_0 = \rho_0 v_0^3$. Using this formula, we
get a similar conversion formula for the energy flux:
\begin{equation}
\label{eq:flux} F = \left( \frac{ \widetilde{F}}{\widetilde{\mu}^2 } \right) \frac{
B_\star^2 (GM_\star)^{1/2} (0.35)^{13/2}}{ R_\star^{1/2} }.
\end{equation}

We use a dimensionless parameter
$\overline{F}_c={\widetilde{F_c}}/{\widetilde{\mu}^2}$ to characterize the flux in the
centers of the spots.

\section{Parameters of Spots}
\label{sec:parameters_spots}

Parameters of the spots for all simulation runs are given in Table \ref{tab:main}.
Simulations were done for five misalignment angles $\Theta=2^\circ, 5^\circ, 10^\circ,
20^\circ, 30^\circ$ and three (dimensionless) periods of the star $P_\star=1.8, 2.4,
2.8$, which correspond to normalized corotation radii of $r_c/R_\star=4.3, 5.1, 5.7$.
In each set of runs with the same parameters $\Theta$ and $P_\star$, we varied the
dimensionless magnetic moment of the star $\widetilde{\mu}$ and observed that the
magnetospheric radius $r_m$ increases with $\widetilde{\mu}$. The magnetospheric
radius is an important parameter because the funnel flow starts approximately at
$r_m$. In this paper, we use the dimensionless ratio $r_m/R_\star$, so that results
can be applied to different types of stars.

Table \ref{tab:main} shows the polar and azimuthal positions of the spots ($\theta_0$,
$\phi_0$), their polar and azimuthal widths ($\theta_1$, $\phi_1$) and the ratio
 $\widetilde{F}_c/\widetilde{\mu}^2$, which characterizes the dimensionless energy flux in
the centers of the spots. If parameters $\Theta$, $r_c/R_\star$ and $r_m/R_\star$ are
known, then Table \ref{tab:main} provides the positions and shapes of the spots.

\begin{figure*}
\centering
\includegraphics[width=17cm]{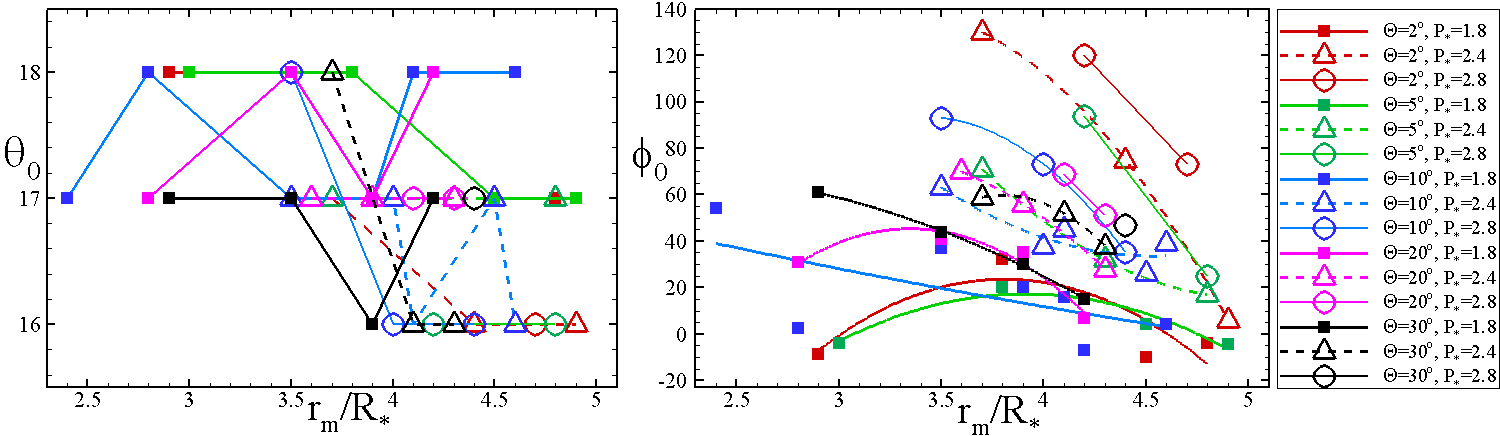}
\caption{\textit{Left panel:} The polar coordinates of the spots position $\theta_0$
in all simulation runs as a function of the magnetospheric radius normalized to the
stellar radius, $r_m/R_\star$, for different misalignment angles: $\Theta=2^\circ$
(red color), $\Theta=5^\circ$ (green color), $\Theta=10^\circ$ (blue color),
$\Theta=20^\circ$ (pink color), and $\Theta=30^\circ$ (black color), and for different
periods of the star: $P_\star=1.8$ (thick solid lines, squares), $P_\star=2.4$ (dashed
lines, triangles), and $P_\star=2.8$ (thin solid lines, circles). The symbols show
results of individual simulation runs, while the lines show interpolation between runs
performed for the same parameters $\Theta$ and $P_\star$.} \label{theta0-phi0}
\end{figure*}

\begin{figure*}
\centering
\includegraphics[width=17cm]{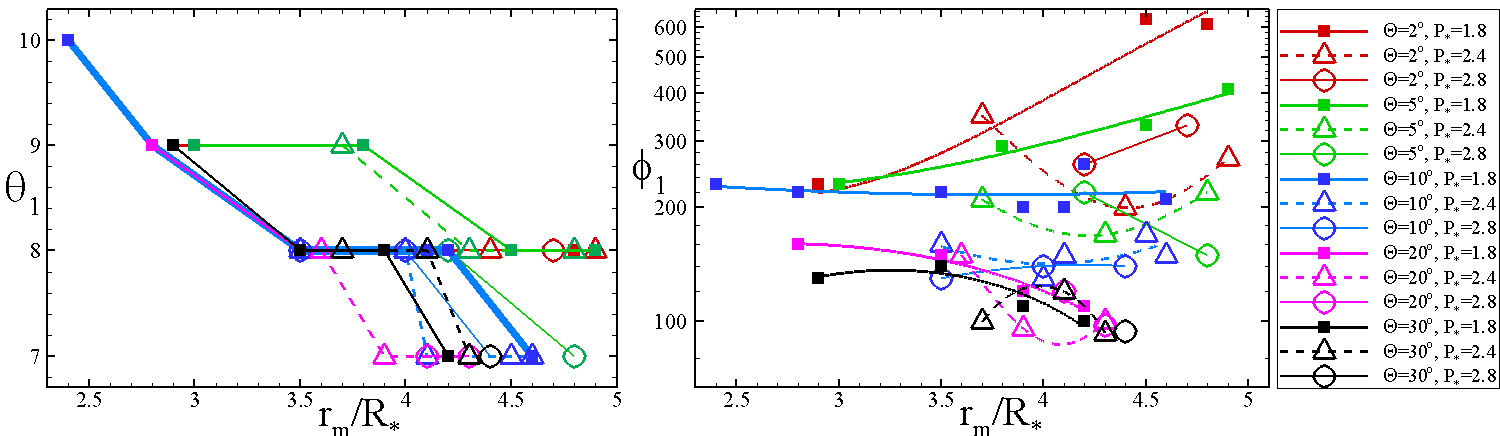}
\caption{Width of the spots in polar ($\theta_1$, see left panel) and azimuthal
($\phi_1$, see right panel) directions as a function of the magnetospheric radius
normalized to the stellar radius, $r_m/R_\star$. Symbols and lines are the same as is
in Fig. \ref{theta0-phi0}.} \label{theta1-phi1}
\end{figure*}

\begin{figure}
\centering
\includegraphics[width=8.5cm]{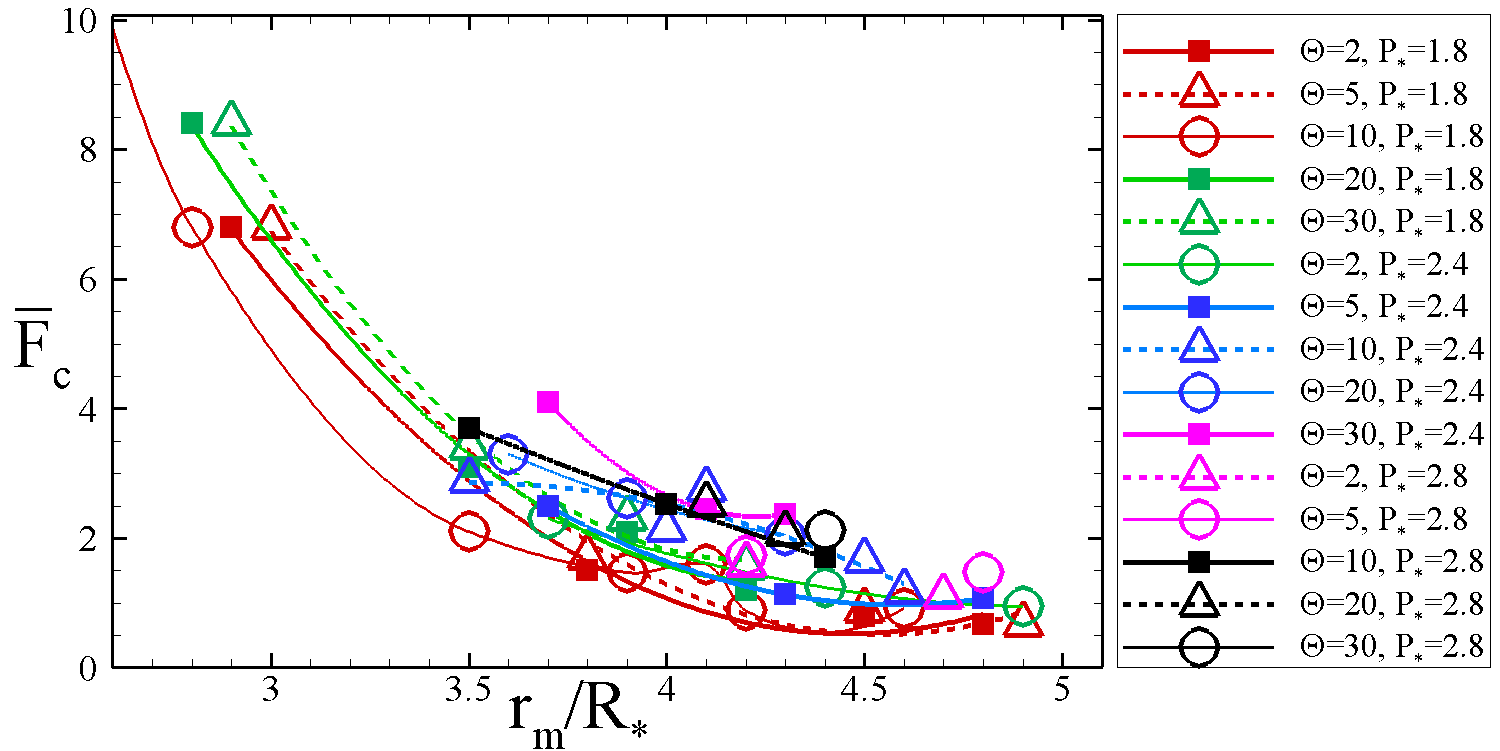}
\caption{Dimensionless function $\overline{F}_c=\widetilde{F}_c/\widetilde{\mu}^2$
(which determines the  energy flux in the centers of the spots) shown for all
simulation runs.  Symbols and lines are the same as is in Fig. \ref{theta0-phi0}.}
\label{F0}
\end{figure}

Figures \ref{theta0-phi0}, \ref{theta1-phi1} and \ref{F0} show parameters of the spots
for all simulation runs.
 Fig. \ref{theta0-phi0} shows the polar
and azimuthal  positions of the spots. Fig. \ref{theta1-phi1} shows the polar and
azimuthal widths of the spots. Fig. \ref{F0} shows dimensionless energy flux in the
center of the spots. A set of runs with the same $\Theta$ is marked by the same color.
A set of runs with the same $P_\star$ is marked by the same symbols. The lines show
interpolations between points corresponding to sets of runs with the same $\Theta$ and
$P_\star$. These lines helped us derive important dependencies discussed in the paper
(see Sec. \ref{sec:results}).

\begin{table*}
\medskip
\noindent\hskip-1.0cm{\bf \Large  $\Theta=2^\circ$}
\begin{tabular}{|c|c||c|c|c|c|c|c|c|c|c|} \hline

$ P_\star$  & $r_m/R_\star\rightarrow$  & 2.9   & 3.7   & 3.8   & 4.2   & 4.4   & 4.5   & 4.7   & 4.8   & 4.9   \\
($r_c/R_\star$)    &     &  &       &       &       &       &       &       &       \\
\hline \hline 1.8
(4.3)     &  $\theta_0,\phi_0$   &  18,~-8.9     &       &  18,~32       &       &       &  17,~-9.8     &       &  17,~-3.9     &  \\
          &  $\theta_1, \phi_1$  &  9,~230       &       &  9,~290       &       &       &  8,~630       &       &  8,~610       &  \\
          &  $\overline{F}_c$  &  1.7        &       &  1.5  &       &       &  1.8  &       &  2.7  &  \\
\hline 2.4
(5.1)       &  $\theta_0,\phi_0$ &       &  17,~130      &       &       &  16,~75       &       &       &       &  16,~6 \\
            &  $\theta_1, \phi_1$ &       &  8,~350       &       &       &  8,~200       &       &       &       &  8,~270 \\
            &  $\overline{F}_c$   &       &  2.3  &       &       &  2.8  &       &       &       &  3.8 \\
\hline 2.8
(5.7)      &  $\theta_0,\phi_0$  &       &       &       &  16,~120      &       &       &  16,~73       &       &  \\
           &  $\theta_1, \phi_1$ &       &       &       &  8,~260       &       &       &  8,~330       &       &  \\
           &  $\overline{F}_c$ &       &       &       &  3.6  &       &       &  4.4  &       &  \\
\hline
\end{tabular}

\medskip
\noindent\hskip-2.7cm{\bf\Large $\Theta=5^\circ$}
\begin{tabular}{|c|c||c|c|c|lc|c|c|c|c|}

\hline
$P_\star$    & $r_m/R_\star \rightarrow$     & 3     & 3.7   & 3.8   & 4.2   & 4.3   & 4.5   & 4.8   & 4.9   \\
($r_c/R_\star$)    &     &  &       &       &       &       &       &       &       \\
\hline \hline 1.8
(4.3)     &  $\theta_0,\phi_0$    &  18,~-3.9     &       &  18,~20       &       &       &  17,~4.4      &       &  17,~-4.6 \\
          &  $\theta_1, \phi_1$   &  9,~230       &       &  9,~290       &       &       &  8,~330       &       &  8,~410 \\
          &  $\overline{F}_c$  &  1.7  &       &  1.7  &       &       &  2    &       &  2.7 \\
\hline 2.4
(5.1)    &  $\theta_0,\phi_0$  &       &  17,~71       &       &       &  17,~32       &       &  17,~17       &  \\
         &       &  9,~210       &       &       &  8,~170       &       &  8,~220       &  \\
          &  $\overline{F}_c$   &       &  2.5  &       &       &  3.8  &       &  4.3  &  \\
\hline 2.8
(5.7)     &  $\theta_0,\phi_0$  &       &       &       &  16,~94       &       &       &  16,~25       &  \\
          &  $\theta_1, \phi_1$   &       &       &       &  8,~220       &       &       &  7,~150       &  \\
          &  $\overline{F}_c$   &       &       &       &  3.9  &       &       &  5.9  &  \\
\hline
\end{tabular}

\medskip
\noindent\hskip-0.7cm{\bf\Large $\Theta=10^\circ$}
\begin{tabular}{|c|c||c|c|c|c|c|c|c|c|c|c|}

\hline
$P_\star$    & $r_m/R_\star \rightarrow$        & 2.4   & 2.8   & 3.5   & 3.9   & 4     & 4.1   & 4.2   & 4.4   & 4.5   & 4.6   \\
($r_c/R_\star$)    &     &  &       &       &       &       &       &       &       \\
\hline \hline 1.8
(4.3)     &  $\theta_0,\phi_0$ &  17,~54       &  18,~2.3      &  17,~37       &  17,~20       &       &  18,~16       &  18,~-7       &       &       &  18,~4 \\
          &  $\theta_1, \phi_1$   &  10,~230      &  9,~220       &  8,~220       &  8,~200       &       &  8,~200       &  8,~260       &       &       &  7,~210 \\
          &  $\overline{F}_c$  &  1.3  &  1.7  &  2.1  &  3.3  &       &  2.5  &  2.7  &       &       &  3.7 \\
\hline 2.4
(5.1)     &  $\theta_0,\phi_0$ &       &       &  17,~63       &       &  17,~38       &  16,~45       &       &       &  17,~26       &  16,~39 \\
          &  $\theta_1, \phi_1$   &       &       &  8,~160       &       &  8,~130       &  7,~150       &       &       &  7,~170       &  7,~150 \\
          & $\overline{F}_c$     &       &       &  2.9  &       &  4.8  &  4.3  &       &       &  5.1  &  4.7 \\
\hline 2.8
(5.7)    &  $\theta_0,\phi_0$ &       &       &  18,~93       &       &  16,~73       &       &       &  16,~35       &       &  \\
         &  $\theta_1, \phi_1$   &       &       &  8,~130       &       &  8,~140       &       &       &  7,~140       &       &  \\
         &  $\overline{F}_c$ &       &       &  3.7  &       &  5.7  &       &       &  6.8  &       &  \\
\hline

\end{tabular}

\medskip
\noindent\hskip-4.5cm{\bf\Large $\Theta=20^\circ$}
\begin{tabular}{|c|c||c|c|c|c|c|c|c|}
\hline
$P_\star$    & $r_m/R_\star \rightarrow$        & 2.8   & 3.5   & 3.6   & 3.9   & 4.1   & 4.2   & 4.3   \\
($r_c/R_\star$)    &     &  &       &       &       &       &       &           \\
\hline \hline 1.8
(4.3)      &  $\theta_0,\phi_0$  &  17,~31       &  18,~41       &       &  17,~35       &       &  18,~6.8      &  \\
           &  $\theta_1, \phi_1$  &  9,~160       &  8,~150       &       &  8,~120       &       &  7,~110       &  \\
           &  $\overline{F}_c$  &  2.1  &  3.1  &       &  4.7  &       &  4.8  &  \\
\hline 2.4
(5.1)     &  $\theta_0,\phi_0$  &       &       &  17,~70       &  17,~56       &       &       &  17,~28 \\
          &  $\theta_1, \phi_1$   &       &       &  8,~150       &  7,~96        &       &       &  7,~100 \\
          &  $\overline{F}_c$  &       &       &  3.3  &  5.9  &       &       &  8.2 \\
\hline 2.8
(5.7)    &  $\theta_0,\phi_0$ &       &       &       &       &  17,~69       &       &  17,~51 \\
         &  $\theta_1, \phi_1$   &       &       &       &       &  7,~120       &       &  7,~98 \\
         &  $\overline{F}_c$  &       &       &       &       &  5.7  &       &  8.3 \\
\hline
\end{tabular}

\medskip
\noindent\hskip-3.4cm{\bf\Large $\Theta=30^\circ$}
\begin{tabular}{|c|c||c|c|c|c|c|c|c|c|}
\hline
$P_\star$    & $r_m/R_\star \rightarrow$        & 2.9   & 3.5   & 3.7   & 3.9   & 4.1   & 4.2   & 4.3   & 4.4   \\
($r_c/R_\star$)    &     &  &       &       &       &       &       &       &       \\
\hline \hline 1.8
(4.3)    &  $\theta_0,\phi_0$  &  17,~61       &  17,~44       &       &  16,~30       &       &  17,~15       &       &  \\
         &  $\theta_1, \phi_1$   &  9,~130       &  8,~140       &       &  8,~110       &       &  7,~100       &       &  \\
         &  $\overline{F}_c$   &  2.1  &  3.4  &       &  5.2  &       &  6.2  &       &  \\
\hline 2.4
(5.1)   &  $\theta_0,\phi_0$   &       &       &  18,~59       &       &  16,~52       &       &  16,~38       &  \\
        &  $\theta_1, \phi_1$    &       &       &  8,~100       &       &  8,~120       &       &  7,~93        &  \\
        &  $\overline{F}_c$   &       &       &  4.1  &       &  5.5  &       &  9.5  &  \\
\hline 2.8
(5.7)   &  $\theta_0,\phi_0$   &       &       &       &       &       &       &       &  17,~47 \\
        &  $\theta_1, \phi_1$    &       &       &       &       &       &       &       &  7,~94 \\
        &  $\overline{F}_c$   &       &       &       &       &       &       &       &  8.5 \\
\hline

\end{tabular}
\caption{\normalsize Parameters of the spots for different dipole misalignment angles
 $\Theta$, three periods of the star, $P=1.8, 2, 2.4$, and different magnetospheric
radii $r_m/R_\star$. Polar and azimuthal angles ($\theta_0$ and $\phi_0$,
respectively) determine the polar and azimuthal positions of the spots. Polar and
azimuthal angles ($\theta_1$ and $\phi_1$, respectively) determine the size of the
spots. The dimensionless function $\overline{F}_c=\widetilde{F}_c/\widetilde{\mu}^2$
determines the energy flux in the centers of the spots.}\label{tab:main}
\end{table*}

\end{document}